\documentclass[useAMS,usenatbib]{mn2e}   
\usepackage{amssymb}
\usepackage{amsmath}
\usepackage{amstext}
\usepackage{deluxetable}
\usepackage{epsfig}
\usepackage{epstopdf}
\usepackage{graphicx}
\usepackage{color}
\usepackage{amsmath}

\newcommand{\HI}{H\,{\sc i}}
\newcommand{\HII}{H\,{\sc ii}}
\newcommand{\NII}{[N\,{\sc ii}]}
\newcommand{\SII}{[S\,{\sc ii}]}
\newcommand{\OII}{[O\,{\sc ii}]}
\newcommand{\OIII}{[O\,{\sc iii}]}

\newcommand{\Ha}{H$\alpha$}
\newcommand{\Hb}{H$\beta$}
\newcommand{\Hg}{H$\gamma$}

\newcommand{\Msun}{~M$_{\sun}$}

\newcommand{\abox}{12+log(O/H)}

\def\Rv{$R_V$}

\newcommand{\Te}{$T_{\rm e}$}

\newcommand{\Ne}{$n_{\rm e}$}

\newcommand{\chb}{$c$(H$\beta$)}
\newcommand{\Wabs}{$W_{abs}$}
\newcommand{\wabs}{$W_{abs}$}

\newcommand{\OHb}{[\ion{O}{iii}]\,$\lambda$5007/H$\beta$}
\newcommand{\NHa}{[\ion{N}{ii}]\,$\lambda$6583/H$\alpha$}

\DeclareRobustCommand{\ion}[2]{%
\relax\ifmmode
\ifx\testbx\f@series
{\mathbf{#1\,\mathsc{#2}}}\else
{\mathrm{#1\,\mathsc{#2}}}\fi
\else\textup{#1\,{\mdseries\textsc{#2}}}%
\fi}

\title[O/H on int-z SN host galaxies]{Elemental gas-phase abundances of intermediate redshift type Ia supernova star-forming host galaxies}

\author[Moreno-Raya et al.] {M.E. Moreno-Raya$^{1}$,
L. Galbany$^{2}$\thanks{E-mail: llgalbany@pitt.edu}, 
{\'A}.R. L{\'o}pez-S{\'a}nchez$^{3,4}$,
M. Moll{\'a}$^{5}$,
\newauthor
S. Gonz\'alez-Gait\'an$^{6}$,
J.M. V{\'i}lchez$^{7}$,
A. Carnero$^{8,9}$
\\
$^1$Centro Astron\'omico Hispano-Alem\'an(CSIC-MPG),
Observatorio Astron\'omico de Calar Alto, Sierra de los Filabres, E-04550 G\'ergal, Almer\'ia, Spain\\
$^2$PITT PACC, Department of Physics and Astronomy, University of Pittsburgh, Pittsburgh, PA 15260, USA\\
$^3$Australian Astronomical Observatory, 105 Delhi Rd, North Ryde, NSW 2113, Australia\\ 
$^4$Department of Physics and Astronomy, Macquarie University, NSW 2109, Australia\\ 
$^5$Departamento de Investigaci{\'o}n B{\'a}sica, CIEMAT, Avda. Complutense 40, 28040, Madrid, Spain\\
$^6$CENTRA, Instituto Superior T\'ecnico, Univerisdade de Lisboa, Portugal\\
$^7$Instituto de Astrof{\'i}sica de Andaluc{\'i}a-CSIC, Apdo. 3004, 18008, Granada, Spain \\ 
$^8$Laborat{\'o}rio Interinstitucional de e-Astronomia - LIneA, Rua Gal. Jos\'e Cristino 77, Rio de Janeiro, RJ 20921-400, Brazil \\ 
$^9$Observat{\'o}rio Nacional, Rua Gal. Jos{\'e} Cristino 77, Rio de Janeiro, RJ 20921-400, Brazil \\
}

\date{Received date: \today; accepted date: ***}

\begin{document}

\maketitle


\begin{abstract}
The maximum luminosity of type Ia supernovae (SNe~Ia) depends
on the oxygen abundance of the regions of the host galaxies where they explode. This metallicity dependence reduces the dispersion in the {\it Hubble diagram} (HD) when included with the traditional two-parameter calibration of SN~Ia light-curve (LC) parameters and absolute magnitude. In this work, we use empirical calibrations to carefully estimate the oxygen abundance of galaxies hosting SNe~Ia from the SDSS-II/SNe Survey at intermediate redshift, by measuring their emission line intensities. We also derive electronic temperature with the direct method for a small fraction of objects for consistency. We find a trend of decreasing oxygen abundance with increasing redshift for the most massive galaxies. Moreover, we study the dependence of the HD residuals (HR) with galaxy oxygen abundance obtaining a correlation in line with those found in other works. In particular, the HR {\sl vs} oxygen abundance shows a slope of -0.186$\pm$0.123\,mag\,dex$^{-1}$ (1.52$\sigma$), in good agreement with theoretical expectations. This implies smaller distance modulii after corrections for SNe~Ia in metal-rich galaxies. Based on our previous results on local SNe Ia, we propose this dependence to be due to the lower luminosity of the SNe~Ia produced in more metal-rich environments.
\end{abstract}

 
\section{Introduction} \label{Section1}

Type Ia supernovae (SNe~Ia) are thermonuclear explosions of carbon-oxygen white dwarfs (WD) in binary systems. 
In the traditional picture, the WD progenitor increases its mass due to accretion from the companion star until reaching the Chandrasekhar mass limit ($\sim1.44$\,\Msun) when the degenerate electron pressure no longer supports its weight. However, there is evidence that question this picture from several aspects: the exact nature of the companion star (another WD or a main sequence star), the actual mass of the progenitor before explosion (sub- or super- Chandrasekhar), or the exact explosion mechanism (deflagration or detonation) is still under debate (see e.g. \citealt{2014ARA&A..52..107M,2013ARA&A..51..457N} ). Anyway, since all explosions in principle occur from WD with similar (although not exact) conditions, it is expected that SNe~Ia show similar luminosities. 

SN~Ia light-curves can be {\sl standardized} by empirical correlations between SNe~Ia peak brightness and their light-curve (LC) width \citep{1993ApJ...413L.105P,1999AJ....118.1766P} and colour \citep{1996ApJ...473...88R}. Several empirical techniques \citep{2004ApJ...613L..21B, 2005A&A...443..781G, 2006ApJ...647..501P, 2007ApJ...659..122J, 2007A&A...466...11G,2011AJ....141...19B} have exploited these correlations and found standard absolute peak magnitudes with a dispersion of 0.10-0.15\,mag, which corresponds to a precision of 5-7$\%$ in the determination of distances \citep{2014A&A...568A..22B,2017arXiv171000845S}. Both the quantity and quality of supernova observations have increased, and limitations on the homogeneity of SNe~Ia have become apparent \citep{1996ApJ...473...88R,2006ApJ...648..868S}. If these inhomogeneities are not accounted for by LC width and color corrections, these variations may introduce systematic errors in the determination of cosmological parameters, preventing reducing further the uncertainties. 

One plausible source of inhomogeneity is the dependence of supernova properties on host galaxy characteristics after other LC corrections. The average properties of host galaxies evolve with redshift, thus any such dependence not included in the standardization techniques will have an effect on the cosmological parameter determination. During the last decade, there have been many studies illustrating the dependence of SNe~Ia light-curve parameters on global characteristics of their hosts \citep{2006ApJ...648..868S, 2008ApJ...685..752G, 2009ApJ...691..661H, 2009ApJ...700..331H, 2010ApJ...715..743K, 2010MNRAS.406..782S, 2010ApJ...722..566L, 2011ApJ...743..172D, 2011ApJ...740...92G, 2011ApJ...734...42N, 2011ApJ...737..102S, 2012ApJ...755..125G, 2013MNRAS.435.1680J}. Some of these latter studies found additionally that SNe~Ia are systematically brighter in massive host galaxies {\bf after LC shape and color corrections}. Through the mass-metallicity relation \cite{2004ApJ...613..898T}, this would lead to a correlation between SN Ia brightness and the metallicity of their host galaxies: metal-richer galaxies host brighter SNe~Ia {\bf after corrections} with a difference of $\sim ~0.10$\,mag, the same order of magnitude of the uncertainties in the standard absolute magnitude. However, the cause of these correlations is not well understood.

In fact, a dependence of the SN Ia peak luminosity with metallicity of the binary system was theoretically predicted: assuming that the mass of the progenitor WD at explosion is fixed, the SN Ia peak magnitude depends mainly on the total quantity of $^{56}$Ni, synthesized during the explosion. \citet{2003ApJ...590L..83T} showed that the magnitude in the LC maximum depends on the WD progenitor chemical abundance of elements C, N, O and $^{56}$Fe: a difference of a factor of 3 in the metallicity may vary the mass of $^{56}$Ni ejected during the explosion in a 25\%.  More recently, \citet{kasen09} analyzed explosion models where asymmetries were taken into account and they found a weak dependence of the SN Ia luminosity on abundances as C and O, claiming that without correcting for this effect, the cosmological distances of these objects may be overestimated by a 2\%. \citet{2010ApJ...711L..66B} also modeled a series of explosions of SNe~Ia finding a non-linear relation between the synthesized mass $M(\rm^{56}Ni)$ and the metallicity $Z$ of the progenitor binary system (see their Fig.1), stronger than the previous results. Summarizing, the luminosity of SNe~Ia may depend crucially on the abundance of elements within the progenitor system, being brighter when $Z$ is lower than for solar abundances. 

Taking all these into account, we have carried out an observational project to further investigate the existence and extent of this relationship. Our final aim is to perform a systematic and careful study of the possible metallicity dependence of the luminosity of SNe~Ia, and to analyze if the dispersion in the distance modulus is reduced when this parameter is taken into account. In order to reach our objective, we planned to determine the metallicity of the SN~Ia progenitor systems using the oxygen abundance of the host galaxies. To do that, we have divided our study in three parts, corresponding to three redshift ranges: 
\begin{enumerate}
\item The local Universe, $z < 0.03$, whose results have been presented  in \citet[hereafter Paper I and Paper II, respectively]{2016MNRAS.462.1281M,2016ApJ...818L..19M}.
\item The intermediate redshift range, $0.03<z<0.5$, for which we analyze a sample of $\sim$400 SN~Ia host galaxies estimating their metallicities and investigating their dependence on SN~Ia LC parameters. The present work is devoted to this part of the project.
\item Finally, the high redshift range, $z > 0.5$, where cosmological models diverge. This will be presented in a future work.
\end{enumerate}

In Paper I and Paper II, we have presented long-slit spectroscopy of 28 SN~Ia host galaxies in the local Universe (z$<$0.03) for which distances had been previously derived using methods independent of SNe~Ia. We estimated the gas-phase oxygen abundance in the region where each SN Ia exploded from the emission lines observed in their optical spectra. We have demonstrated that there is a correlation between SN~Ia absolute magnitudes and oxygen abundances at the explosion sites, in the sense that SNe Ia tend to be brighter at metal-poorer locations. This relation may explain why after standardizing SN Ia luminosities, brighter SNe~Ia are usually found in metal-richer or more massive galaxies, implying that the standard calibration tends to overestimate the luminosities of these objects. In this way, if the metallicity dependence is neglected, the standardized luminosity of a SN~Ia in a metal-rich environment would be higher than the actual value, thus showing a residual to the best cosmology of the same order than  actually observed. Furthermore, we estimated from our local sample that when this oxygen abundance dependence is included, the dispersion in $M_{B}$ is reduced by $\sim 5\%$.

In this work we will concentrate on point (ii) estimate oxygen abundances for a sample of intermediate redshift SN~Ia host galaxies. \citet{2016MNRAS.457.3470C} presented a systematic study of the relationships between the SNe~Ia properties and the host galaxies characteristics for a sample of 581 SNe~Ia from the SDSS-II SN survey, which covers a redshift range of $0.05 < z < 0.55$. They find a slight correlation between HR and host galaxy metallicity, and also a stronger correlation with the host galaxy stellar mass. Moreover, they studied possible effects of these dependences on cosmological parameters, allowing a new free parameter in the fit of data to the models in the HD, resulting in a shift over the parameter $w$ of the equation of state towards a more negative value, $w\sim -1.15^{+0.123}_{-0.121}$, compared with the value of $w \sim -0.970$ obtained without any dependence on mass or metallicity (a 18\% of difference). An important caveat of this work is that they also use data from passive galaxies, for which there are no emission lines with sufficient S/N, obtaining however some upper limits from the continuum flux in the regions where the lines are expected. In this way they obtain two different zones in the HR as a function of oxygen, where these upper limit abundances showing values larger that $12+log(O/H) > 8.8-8.9$~dex, may not be considered valid. Even so, they find a correlation with a slope -0.155\,mag\,dex$^{-1}$, that implies a difference of $\sim 0.150$\,mag between objects with $12+log(O/H)=8.0$\,dex and 9.0\,dex, the most metal rich, after LC shape and color correction, being brighter. In addition, the emission lines used in that work are not directly measured but taken from tables in \citet{2013MNRAS.431.1383T}. Due to the high number of spectra, these line intensities were obtained by automatic methods which are not always precise enough as already shown in \citet{2016ApJ...821..115W}, who instead measured their own intensities, optimizing the method to subtract the continuum component, for which they needed to reduce the redshift range to $z<0.3$. They measured H$_{\alpha}$ and [NII] simultaneously, fixing the velocity, too. Thus, they improve the measurements of the emission lines despite the reduction of redshift range, and, in consequence, the number of objects to 144. 

In this work we estimate the oxygen abundances for a sample of intermediate redshift SN Ia host galaxies. Our final aim is to check if the SNe luminosities maintain the dependence on oxygen abundances as we found previously in the local Universe, and if the HR dependence on metallicity vanishes when a new parameter $\gamma$ is considered. We carefully measured all the emission lines determining the stellar continuum and the Balmer absorption when possible. The selection of the host galaxy sample is presented in Section 2. The analysis of the galaxy spectra and the determination of oxygen abundances is detailed in Section 3, where the estimation of oxygen abundances from intensity ratios by the different empirical calibration methods is discussed, In Section 4 we discuss our results, and our conclusions are given in Section 5.


\section{Sample Selection}  \label{Section2}

\subsection{The SDSS-II Supernova Survey}

The Sloan Digital Sky Survey-II Supernova Survey (SDSS-II/SNe) \citep{2008AJ....135..338F} has identified and measured LCs for intermediate redshift (0.01$<$ $z$ $<$ 0.45) SNe~Ia during the three fall seasons of operation from 2005 to 2007, using the dedicated SDSS 2.5m telescope at Apache Point Observatory \citep{1998AJ....116.3040G,2006AJ....131.2332G}. These SNe are all located in Stripe 82, a 2.5 degree wide region along the Celestial Equator from roughly -50 $<$ $RA$ $<$ 59 in the Southern Galactic hemisphere \citep{2002AJ....123..485S}.  This effort resulted in 536 SNe~Ia confirmed spectroscopically (sp-Ia), and 914 transients photometrically classified as SNe~Ia (ph-Ia) from their LCs \citep{2014arXiv1401.3317S} and using the spectroscopic redshift of the host galaxy either measured previously by the SDSS Legacy Survey \citep{2000AJ....120.1579Y} or by the BOSS Survey \citep{2013AJ....145...10D}. In addition, we add to this sample 16 SNe~Ia from the pilot SDSS-II/SNe survey performed in 2004, doing a total of 1466 host SNe~Ia galaxies. We finally select those galaxies with optical spectra publicly available in the SDSS Data Release 12 (DR12, \citealt{2015ApJS..219...12A}). From the initial set of 1466 galaxies, 1128 (344 sp-Ia and 784 ph-Ia) pass the above criterion.

\subsection{The Union2.1 compilation}

We repeat the procedure with the SNe in the Union 2.1 compilation \citep{2012ApJ...746...85S}, which contains SNe~Ia from several previous works with good quality light-curves and available data to use them for cosmological purposes. From the initial set of 580 SNe~Ia, we find 60 with their host galaxy  spectra publicly available in SDSS DR12. These 60 SNe are at the lowest redshift range of the Union2.1 (hereafter U2.1) compilation, between 0.015 and 0.17, with 20 of them (a 30\%) with $z < 0.05$. The original publication sources are described in \cite{2012ApJ...746...85S}.

\begin{figure}
\centering
\includegraphics[trim=1cm 0.3cm 0.5cm 0.9cm,clip=True,width=\columnwidth]{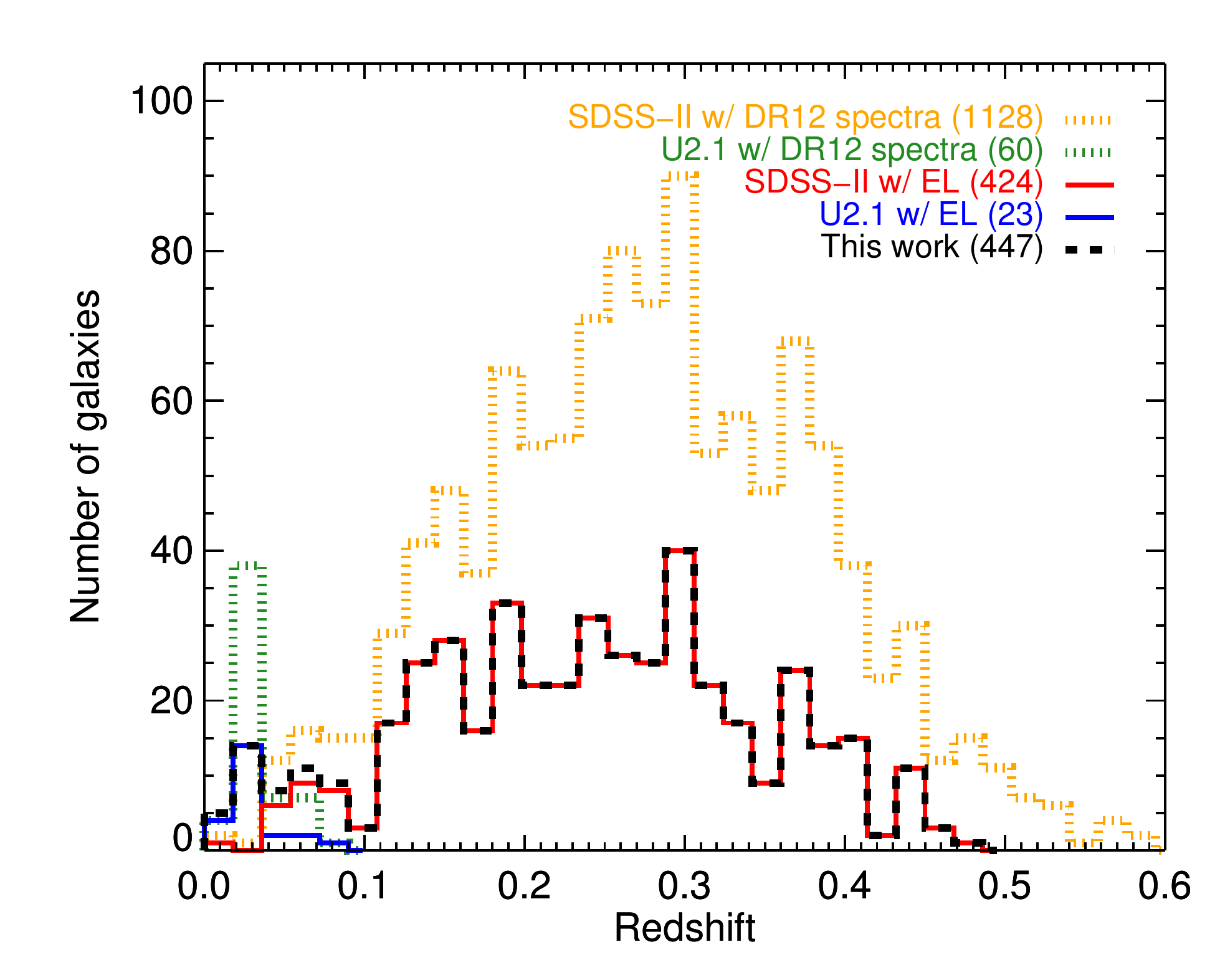}     
\caption{Histogram showing redshift distribution of our spectroscopic sample of SNe~Ia host galaxies from SDSS DR12. In red (SDSS) and blue (Union 2.1) our analyzed sample extend above all redshifts. In green the initial complete sample (60) of Union2.1 with spectra, and in orange, the 1128 SDSS SNe~Ia hosts galaxies with spectra. The black dashed line denotes the final distribution of 447 objects from both sources used in this work.}
\label{fig:figure01}
\end{figure}

\subsection{Host galaxy spectroscopic sample}

\begin{figure*} 
\centering
\includegraphics[width=\linewidth]{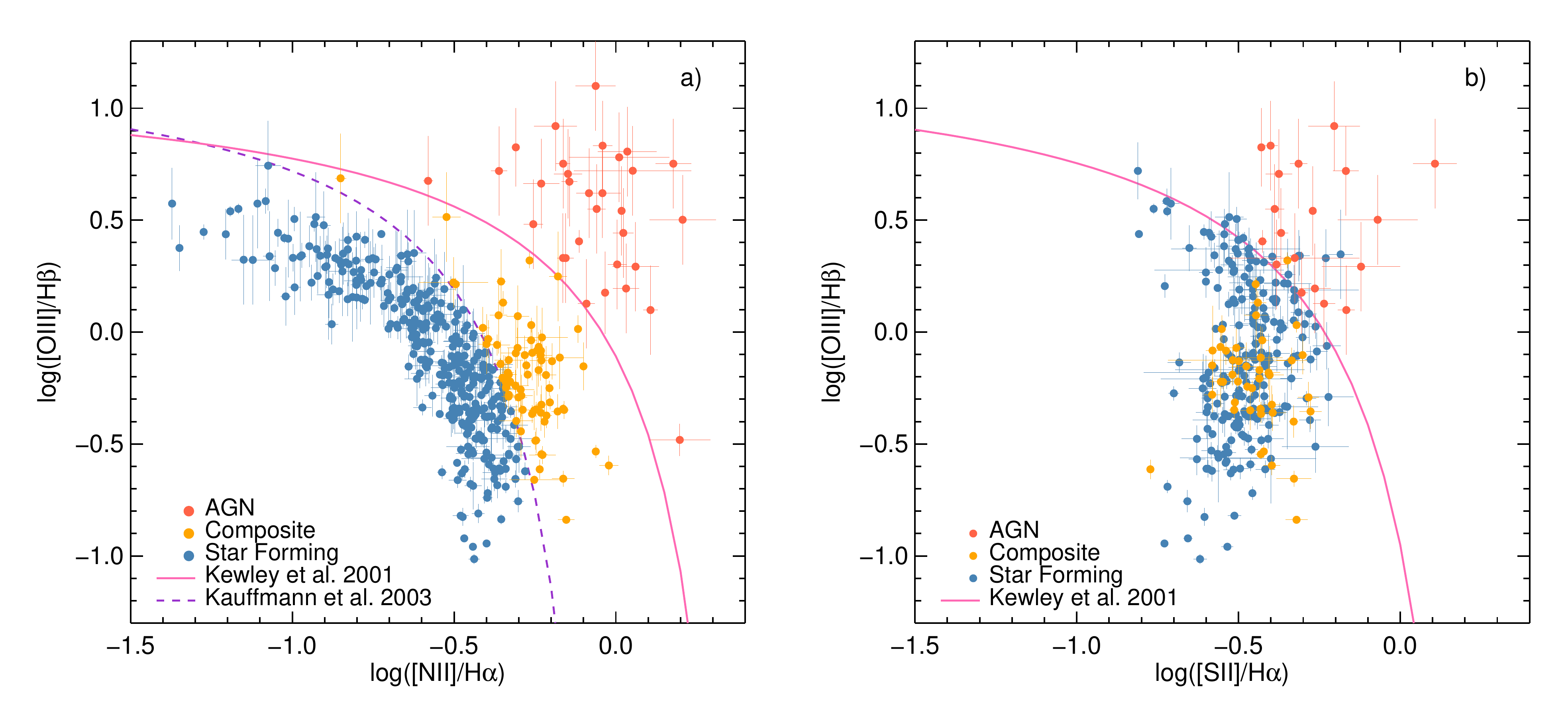}   
\caption{Diagnostic diagrams. Left) Observational flux ratios, \OHb\ {\it vs} \NHa, obtained analyzing our galaxy sample. The theoretical line derived by \citet{2001ApJ...556..121K} that divides the emisson between ionization from massive stars and AGNs is shown in pink. In a purple dashed line we show the empirical line from \citet{2003MNRAS.346.1055K} that divides further the space between pure star-forming and composite regions. Blue circles indicate pure star-forming galaxies, yellow points are the galaxies with composite nature, while galaxies classified as AGN are represented in red. Right)  \OHb\ {\it vs} SII/\Ha, with a similar \citet{2001ApJ...556..121K}'s line dividing AGNs from composite and star forming objects.}
\label{fig:figure02}
\end{figure*}

The spectral SN~Ia host galaxy sample we use in this work consists, therefore, of 1188 objects: 1128 from the SDSS-II/SNe survey and 60 from the Union 2.1 compilation. The corresponding spectra present different qualities, depending on the presence of sky lines, the SNR ratio or faintness of the lines. In the next section, we detail the analysis performed with these spectra. The wavelength range in the observer frame covers from 3700 to 9000\AA\, so this permits to measure the needed emission lines up to redshift 0.45, where H$\alpha$ and \NII~$\lambda$6583 are located at 9545\AA.

Spectra were obtained with a 2" or 3'' diameter fiber\footnote{While SDSS-I and SDSS-II spectra were observed with a fiber with 3" diameter, BOSS spectra were observed with a 2" diameter fiber.} pointed to the center of each galaxy, so the light captured by the fiber corresponds to different fractions of the galaxy depending on the galaxy projected size in the sky and the redshift. This may imply a difference between the real metallicity in the zone where SNe~Ia exploded and the one estimated which will correspond to different fractions of the galaxy \citep{2016A&A...591A..48G}, as a consequence of the metallicity radial gradients usually shown in galaxies \citep{1999PASP..111..919H,2005MNRAS.358..521M}, but we have not corrected for this effect. A detailed study of aperture effects on using fiber spectra in SN Ia cosmology will be presented elsewhere (Galbany et al. in prep.).

In Figure \ref{fig:figure01} we show the distribution in redshift for the sample: In light and dark grey are the complete sample we
use for SDSS and Union2.1, respectively. In red (SDSS) and purple (Union2.1), we show the subsample for which we measure
emission lines with the adequate signal-to-noise  (see next Section).


\section{Analysis of the host galaxy spectra} 
\label{Section3}

\subsection{Line measurement and nature of the emission\label{diagnostic}}

First, we correct all spectra for Milky Way dust extinction using the dust maps of \citet{2011ApJ...737..103S} retrieved from the NASA/IPAC Infrared Science Archive (IRSA), applying the standard Galactic reddening law with \Rv = 3.1 \citep{1989ApJ...345..245C,1994ApJ...422..158O}. We also shift the spectra to the rest frame wavelengths.

In these 1188 spectra we try to measure seven emission lines: three hydrogen Balmer lines (\Ha, \Hb, \Hg) and the brightest collisional excited lines of heavy elements: \OII$\lambda\lambda$3726,29 (blended), \OIII$\lambda$5007, \NII$\lambda$6583, and \SII~$\lambda\lambda$6716,31. All measurements are performed manually using the {\sc splot} routine of {\sc iraf}\footnote{Image Reduction Analysis Facility, distributed by NOAO which is operated by AURA Inc., under cooperative agreement with NSF}. Line intensities and equivalent widths are measured integrating all the emission between the limits of the line and over a local adjacent continuum. However, because of the faintness of many of the detected emission lines, a detailed inspection of the spectra is always necessary to get a proper estimation of the adjacent continuum and the line flux in these cases. For each emission line (in particular, for the hydrogen Balmer lines, as they are affected by absorptions of the stellar component of the galaxy) we consider several measurements at slightly different continuum levels. For each spectrum, we choose the emission line fluxes which provide the best match and minimize the uncertainties of the derived properties (reddening coefficient and oxygen abundance). Uncertainties in the line fluxes were estimated for each line considering both the $rms$ of the continuum and the width of each emission line. We require galaxies to have lines of \Ha, \Hb, and \NII~$\lambda$6583 detected at greater than 5$\sigma$. We constrain as well the other nebular lines that we make use of to be detected at greater than 3$\sigma$. With these cuts our sample reduces to 424 galaxies from SDSS and 23 from Union2.1 for a total of 447 galaxies to analyze (see Table \ref{lineid} and redshift distribution on Figure \ref{fig:figure01}).

We start by checking the nature of the ionization of the gas within our sample galaxy using the so-called diagnostic diagrams, as firstly proposed by \citet{1981PASP...93....5B} and \citet{1987ApJS...63..295V}.  These diagrams are useful tools to distinguish between pure star-forming regions, Low-Ionization Narrow-Emission Region (LINER) activity, and Active Galactic Nucleus (AGN). By using the observed emission line intensities, we show in Figure \ref{fig:figure02} the typical ratios \OIII~$\lambda$5007/\Hb\ versus \NII~$\lambda$6583/\Ha\ and \OIII~$\lambda$5007/\Hb\ versus (\SII~$\lambda$6716+$\lambda$6731)/\Ha\ diagrams for all the galaxies for which we have a good measurement of these lines. In these diagnostic diagrams, \HII\ regions and starburst galaxies lie within a narrow band.  We plot in the same diagrams the analytical relations given by \citet{2001ApJ...556..121K}, which were derived for starburst galaxies and which represent an upper envelope of positions of star-forming galaxies. In the left panel of Fig.~{\ref{fig:figure02} we also draw the empirical relation between \OIII~$\lambda$5007/\Hb\ and \NII~$\lambda$6583/\Ha\ from \cite{2003MNRAS.346.1055K} derived after analyzing a large data sample of star-forming galaxies from SDSS data. Galaxies below this curve are considered pure star-forming objects. When the gas is excited by shocks, accretion disks, or cooling flows (as in the case of AGNs or LINERs) its position in these diagnostic diagrams is away from the locus of \HII\ regions, and usually well above those curves. It is usually considered that the nature of the ionization of the gas in galaxies located in this region is not due to massive stars, i.e., that these objects are not star-forming galaxies. We plot galaxies above the \citet{2001ApJ...556..121K} curve using a red color and we will not further consider them in our analysis.

\citet{2006MNRAS.372..961K} suggested that those objects between the theoretical line computed by \citet{2001ApJ...556..121K} and the empirical line found by \citet{2003MNRAS.346.1055K} may be ionized by both massive stars and shocks, i.e., to have a composite nature, although \citet{2009MNRAS.398..949P} claimed that objects located in this area may also be pure star-forming galaxies but with a high \NII~content. In any case, we use a yellow color to distinguish these galaxies, which will be also analyzed in this work.

In the right-panel of Fig.~{\ref{fig:figure02}, we show a similar diagnostic diagram using the ratio \SII/\Ha\  where all objects previously classified as AGNs also lie above the theoretical curve. Although we are not able to measure \SII $\lambda\lambda$~6717,31 lines in 38\% of the objects classified as star-forming or composite in the BPT diagram (panel a), and therefore they are not included in the sulfur diagnostic diagram (panel b), we keep these objects in our analysis given their reliable classification in the BPT. From the 447 galaxies analyzed, 340 have been labeled as SF, while 76 have composite nature and 31 have been classified as AGNs, so, we have finally obtained/analyzed a total of 416 galaxy spectra.

\subsection{Reddening correction}

\begin{figure}
\centering
\includegraphics[trim=0.7cm 0cm 0.5cm 0cm, clip=True,width=\columnwidth]{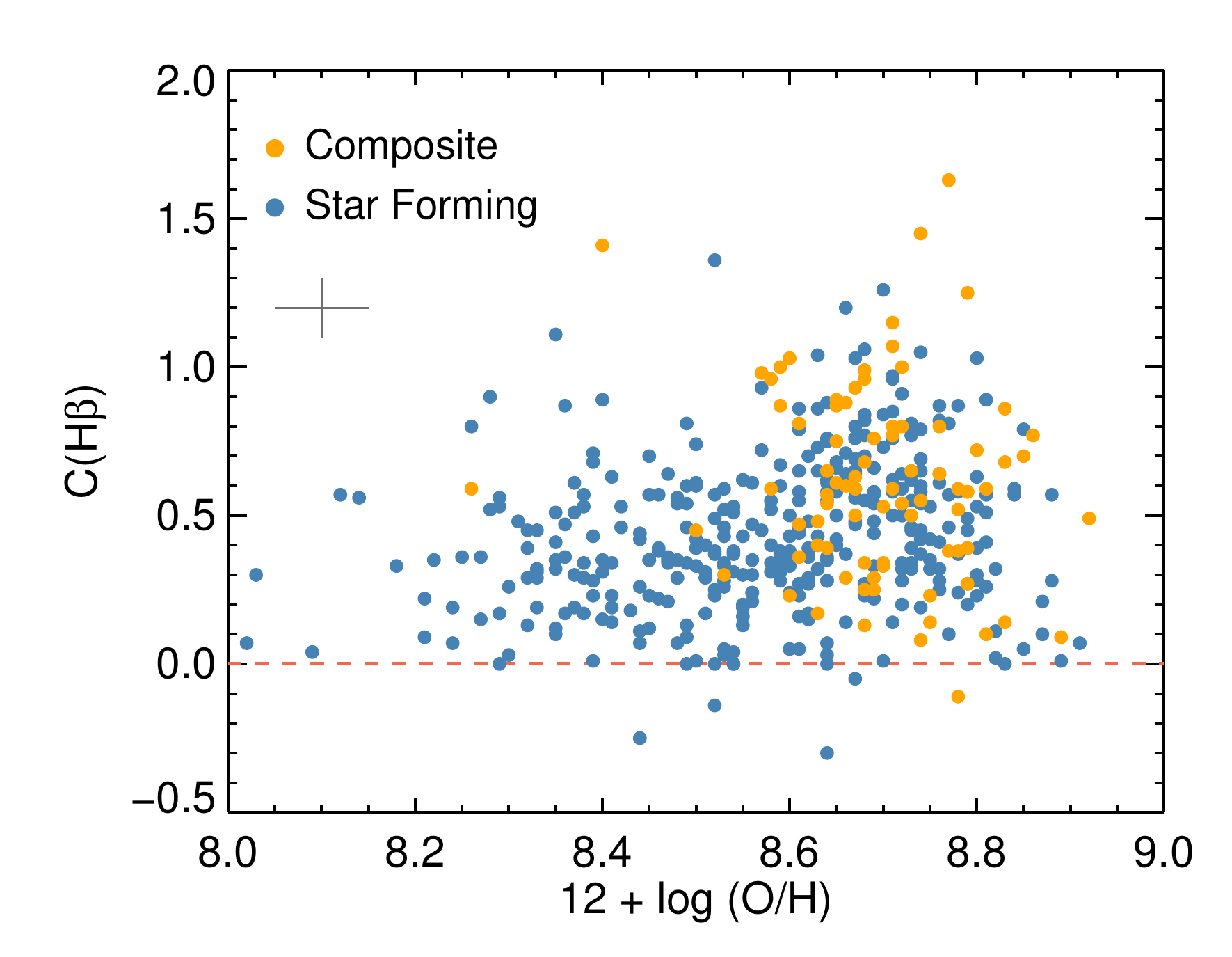}   
\caption{Reddening coefficient c(H$\beta$).  Galaxies with composite AGN/star-forming nature are plotted in yellow, and pure star-forming galaxies in blue. The coefficient c(H$\beta$) obtained from the ratio $I$(\Ha)/$I$(\Hb) as a function of the oxygen abundance as $12+log(O/H)$. The (red) short-dashed line represents the no-reddening line c(H$\beta=0$).}
\label{fig:figure03}
\end{figure}

Emission lines fluxes must be corrected for the host galaxy reddening. This correction is usually made  using the Balmer decrement between the $H\alpha$ and $H\beta$ line fluxes according to: 
\begin{equation}
\label{redeq}
\frac{I(\lambda)}{I(H\beta)} = \frac{F(\lambda)}{F(H\beta)}10^{c(H\beta)[f(\lambda)-f(H\beta)]},
\end{equation}
where $I(\lambda)/I(H\beta)$ is the line intensity flux unaffected by reddening or absorption, $F(\lambda)/F(H\beta)$ is the observed line measured flux, $c(H\beta)$ is the reddening coefficient and $f(\lambda)$ is the reddening curve normalized to \Hb\ using the \citet{1989ApJ...345..245C} law with $R_V=3.1$.

\begin{deluxetable}{c@{\hspace{5pt}}c@{\hspace{5pt}}    cc@{\hspace{5pt}}c@{\hspace{5pt}} cc@{\hspace{5pt}}c@{\hspace{5pt}}  cc@{\hspace{5pt}}c@{\hspace{5pt}}  cc    c c c    ccc } 
\rotate
\tabletypesize{\scriptsize}
\tablecaption{Dereddened line intensity ratios with respect to $I$(\Hb)=100 or  $I$(\Ha)=100. Colons indicate measurements with an uncertainty larger than 40\%. Three dots indicate that the emission line has not been detected. Only some galaxies are included here, the whole table will be given in electronic format. \label{lineid}}
\tablewidth{0pt}
\tablehead{
\colhead{Object} &  
\colhead{Nature} &  
\multicolumn{2}{c}{ $\frac {\rm [O\,II]]\,\lambda3727}{\rm H\beta}$  } & 
\multicolumn{2}{c}{ $\frac {\rm H\gamma}{\rm H\beta}$} & 
\multicolumn{2}{c}{ $\frac {\rm [O\,III]]\,\lambda5007}{\rm H\beta}$} & 
\multicolumn{2}{c}{ $\frac {\rm H\alpha}{\rm H\beta}$} &
\colhead{ $\frac {\rm [N\,II]]\,\lambda6583}{\rm H\alpha}$} & 
\colhead{ $\frac {\rm [S\,II]]\,\lambda6716}{\rm H\alpha}$} & 
\colhead{ $\frac {\rm [S\,II]]\,\lambda6731}{\rm H\alpha}$} &
\colhead{$W$(\Hg)} & 
\colhead{$W$(\Hb)} & 
\colhead{$W$(\Ha)} \\ 
& & $I$ & $I_0$ & $I$ & $I_0$ & $I$ & $I_0$ & $I$ & $I_0$ & $I$ & $I$ & $I$ & [\AA] & [\AA] & [\AA]
}
\startdata
5		&	SF	&	130$\pm$0	&	141$\pm$0	&	...			&	...			&	23$\pm$10	&	22$\pm$11	&	368$\pm$28	&	304$\pm$32	&	43$\pm$3		&	18$\pm$3		&	13$\pm$2		&	...			&	-4.8$\pm$0.5	&	-14.0$\pm$0.3	\\
10		&	SF	&	321$\pm$0	&	418$\pm$0	&	42$\pm$3		&	46$\pm$5		&	262$\pm$4	&	254$\pm$4	&	387$\pm$5	&	285$\pm$5	&	9.5$\pm$0.7	&	18.5$\pm$0.5	&	14.3$\pm$0.5	&	-7.2$\pm$0.6	&	-13.0$\pm$0.2	&	-60.9$\pm$0.3	\\
30		&	SF	&	136$\pm$3	&	256$\pm$7	&	32$\pm$1		&	46$\pm$2		&	42$\pm$1		&	39$\pm$1		&	545$\pm$5	&	299$\pm$4	&	36.6$\pm$0.4	&	16.8$\pm$0.4	&	12.7$\pm$0.4	&	-4.47$\pm$0.18	&	-12.85$\pm$0.18	&	-69.8$\pm$0.2	\\
83		&	SF	&	285$\pm$64	&	319$\pm$86	&	...			&	...			&	69$\pm$15	&	68$\pm$17	&	425$\pm$44	&	294$\pm$42	&	33$\pm$3		&	27$\pm$2		&	21$\pm$2		&	...			&	-3.0$\pm$0.4	&	-10.4$\pm$0.4	\\
128		&	SF	&	346$\pm$0	&	440$\pm$0	&	44$\pm$3		&	46$\pm$4		&	258$\pm$6	&	251$\pm$7	&	376$\pm$9	&	288$\pm$9	&	16.7$\pm$1.5	&	20.5$\pm$1.1	&	11.3$\pm$1.1	&	-7.5$\pm$0.5	&	-16.4$\pm$0.5	&	-73.0$\pm$0.8	\\
133		&	SF	&	96$\pm$0		&	155$\pm$0	&	...			&	...			&	18$\pm$9		&	17$\pm$9		&	549$\pm$22	&	306$\pm$17	&	34.2$\pm$1.1	&	14.5$\pm$1.4	&	10.5$\pm$1.2	&	...			&	-5.5$\pm$0.3	&	-22.3$\pm$0.4	\\
171		&	SF	&	311$\pm$0	&	391$\pm$0	&	...			&	...			&	118:     		&	115:     		&	381$\pm$14	&	290$\pm$15	&	23.0$\pm$1.7	&	21.6$\pm$1.9	&	17.0$\pm$1.3	&	...			&	-9.3$\pm$0.5	&	-35.7$\pm$0.6	\\
... \\
\enddata
\end{deluxetable}

\begin{deluxetable}{c@{\hspace{0pt}} c@{\hspace{0pt}} cc@{\hspace{2pt}}    cc@{\hspace{2pt}} c@{\hspace{0pt}}  cc@{\hspace{2pt}} cc@{\hspace{2pt}}  c@{\hspace{0pt}} c@{\hspace{0pt}}    cc@{\hspace{0pt}}  cccc@{\hspace{0pt}} } 
\rotate
\tabletypesize{\scriptsize}
\tablecaption{Derived properties from the oxygen abundances estimation for galaxies classified as Star Forming or Composite. \label{ohtable}}
\tablewidth{0pt}
\tablehead{
\colhead{Object} &\colhead{Class} & \colhead{$c$(\Hb)} & \colhead{$W_{abs}$} &\multicolumn{10}{c}{ 12 + log(O/H)  } &\colhead{$Te$ \OIII} &\colhead{log $\frac{\rm O^{++}}{\rm O^{+}}$} &\colhead{12+log(N/H) } &\colhead{log(N/O) } \\
\cline{5-14} \\
& & & [\AA] &~P01~&~PT05~&~M13a~&~M13b~&~PP04a~&~PP04b~&~KK04T~&~KDN2O2~&~${\rm OH_{FINAL}}$~}
\startdata
5	&	SF	&	0.11	$\pm$	0.06	&	1.0	$^{a}$		&	8.47	&	8.75	&	8.58	&	8.57	&	8.69	&	8.82	&	8.82	&	8.82	&	8.76	$\pm$	0.04	&	&	5175	&	-0.48	&	8.03	&	-0.68	$\pm$	0.03	\\
10	&	SF	&	0.36	$\pm$	0.01	&	1.0	$\pm$	0.1	&	8.10	&	8.22	&	8.22	&	8.27	&	8.32	&	8.27	&	8.28	&	...	&	8.26	$\pm$	0.05	&	&	11175	&	-0.3	&	6.66	&	-1.43	$\pm$	0.03	\\
30	&	SF	&	0.85	$\pm$	0.01	&	0.3	$\pm$	0.1	&	8.26	&	8.55	&	8.52	&	8.54	&	8.65	&	8.71	&	8.73	&	8.68	&	8.66	$\pm$	0.03	&	&	6475	&	-0.7	&	7.7	&	-0.93	$\pm$	0.03	\\
83	&	SF	&	0.15	$\pm$	0.12	&	1.0	$^{a}$		&	8.08	&	8.33	&	8.46	&	8.53	&	8.63	&	8.63	&	8.61	&	8.61	&	8.58	$\pm$	0.02	&	&	7425	&	-0.64	&	7.61	&	-0.92	$\pm$	0.08	\\
128	&	SF	&	0.32	$\pm$	0.02	&	1.0	$\pm$	0.1	&	8.07	&	8.19	&	8.28	&	8.38	&	8.46	&	8.35	&	8.27	&	...	&	8.33	$\pm$	0.04	&	&	10625	&	-0.32	&	6.95	&	-1.23	$\pm$	0.03	\\
133	&	SF	&	0.63	$\pm$	0.05	&	1.0	$^{a}$		&	8.32	&	8.70	&	8.61	&	8.52	&	8.63	&	8.82	&	8.81	&	8.77	&	8.75	$\pm$	0.04	&	&	5275	&	-0.66	&	7.91	&	-0.8	$\pm$	0.03	\\
... \\
\enddata
\footnotesize
\tablecomments{
P01: \citet{2001A&A...369..594P,2001A&A...369..594P} using $R_{23}$ and $P$; 
PT05: \citet{2005ApJ...631..231P} using $R_{23}$ and $P$; 
M13a: \citet{2013A&A...559A.114M} using the $O3N2$ parameter;
M13b: \citet{2013A&A...559A.114M} using the $N2$ parameter;
PP04a: \citet{2004MNRAS.348L..59P} using a linear fit to the $N2$ parameter; 
PP04b: \citet{2004MNRAS.348L..59P} using the $O3N2$ parameter;
KK04T: \citet{2004ApJ...617..240K} using $R_{23}$ and the ionization parameter defined in that paper, $q_{KK04}$; 
KDN2O2: \citet{2002ApJS..142...35K} using the $N2O2$ parameter (calibration only valid for objects in the high metallicity.}
\end{deluxetable}

To calculate the reddening coefficient, \chb, both $H\alpha$ and $H\beta$ are usually used in Eq.~\ref{redeq}. Sometimes other pairs of \HII\ Balmer lines --e.g., \Hg/\Hb\ or H$\delta$/\Hb-- can be measured too and the reddening coefficient can be determined with higher accuracy \citep[e.g., see][]{2009A&A...508..615L}.  However, in extragalactic objects, the fluxes of nebular Balmer lines are affected by the absorption produced by the underlying stellar population (mainly B and A stars). We here consider that the Balmer lines are indeed affected by underlying stellar absorptions, and have taken into account the absorption in the hydrogen lines, $W_{abs}$ --which we assume is the same for all the Balmer lines-- following:
\begin{equation}
c(H\beta)=\frac{1}{f(\lambda)} \log\Bigg[\frac{\frac{I(\lambda)}{I(H\beta)}\times
\Big(1+\frac{W_{abs}}{W_{H\beta}}\Big)} {\frac{F(\lambda)}{F(H\beta)}\times 
\Big(1+\frac{W_{abs}}{W_{\lambda}}\Big)}\Bigg],
\end{equation}
as introduced by \citet{1993ApJS...85...27M}, where $W_{abs},\ W_{\lambda}$, and $W_{\rm H\beta}$ are the equivalent widths of the underlying stellar absorption, the considered Balmer line, and H$\beta$, respectively.

For some galaxies we are able to observe three \HI\ Balmer lines (\Ha, \Hb, and \Hg). Hence, we derive the best \chb\ and \wabs\ that match the observed fluxes and equivalent widths. In the cases where only two Balmer lines (i.e., \Ha\ and \Hb) are detected, we assume \Wabs=1.0~\AA\ and derive the reddening coefficient using only the \Ha/\Hb\ ratio.

Typical analyses of star-forming galaxies usually consider that the theoretical $I$(\Ha)/$I$(\Hb) intensity ratio is 2.86, following the case B recombination for an electron density of \Ne=100\,cm$^{-2}$ and electron temperature of \Te=10000\,K. However, the theoretical \HI\ Balmer ratios have a dependence on the electron temperature --see \citet{1995MNRAS.272...41S}  and also Appendix~A in \citet{2015MNRAS.450.3381L} --. Actually, this value of 2.86 changes to $I$(\Ha)/$I$(\Hb)=3.01 for \Te$\sim$5000\,K (and reduces for higher \Te values). The values of these ratios are, in turn, related to the metallicity of the ionized gas, in the sense that objects with low (high) electron temperature have high (low) oxygen abundance. We use the prescriptions given by \citet{2015MNRAS.450.3381L} --see their Appendix ~A-- to consider the dependence of the theoretical \HI\ Balmer line ratios on the electronic temperature, assuming the best value to the oxygen abundance provided by the empirical calibrations (see next subsection).

Once we  determine the reddening coefficient, we correct all the emission lines we measure by using Eq.~\ref{redeq}. The resulting line intensities are given in Table~\ref{lineid}, (given in electronic format, we give here a portion as example), which lists for each galaxy (named with the supernova Ia hosted by the galaxy) in column 1, the nature of the emission as a symbol C, SF or A for composite, star-forming or AGN, respectively, in column 2, the corrected emission line intensities as $\frac {\rm [O\,II]\,\lambda3727}{\rm H\beta}$, $\frac {\rm H\gamma}{\rm H\beta}$, $\frac {\rm [O\,III]\,\lambda5007}{\rm H\beta}$ and $\frac {\rm H\alpha}{\rm H\beta}$, in units of $I(H\beta)=100$, with their errors, in columns 3 to 10.  For each line the observed, I($\lambda$), and theoretical, corrected for reddening, intensities, I$_0$($\lambda$), are given. Moreover, the emission line intensities $\frac {\rm [N\,II]\,\lambda6583}{\rm H\alpha}$, $\frac {\rm [S\,II]\,\lambda6716}{\rm H\alpha}$ , and $\frac {\rm [S\,II]\,\lambda6731}{\rm H\alpha}$ in units of $I(H\alpha)=100$, with their errors, are given in columns 11 to 13. The equivalent widths in \AA\ for \Ha, \Hb, and \Hg\ emission lines are in columns 14 to 16, respectively. Colons indicate measurements with an uncertainty larger than 40\%, including their uncertainties.

Figure~\ref{fig:figure03} shows the reddening coefficient c(H$\beta$) that we obtain from the ratio $I$(\Ha)/$I$(\Hb) as a function of the final estimated oxygen abundance (see next section). The red dashed line indicates the case where no reddening appears. In most of the star-forming galaxies the reddening is smaller than 1.0 dex, with an average value of $\sim 0.4$ dex, while the composite objects show higher values reaching 1.5\,dex in some cases. As seen in \citet{2010A&A...517A..85L}, Figure~\ref{fig:figure03} seems to show the trend that the reddening coefficient increases with the oxygen abundance. Table~\ref{ohtable} compiles the \chb\ and \wabs\ values --and their associated errors-- derived for each galaxy. (see next section for details about the table).

\subsection{Chemical Abundances}

\subsubsection{Empirical Methods to estimate abundances}

When the faint auroral lines, such as \OIII~$\lambda$4363 or \NII~$\lambda$5755, are not detectable, the so-called strong emission line (SEL) methods should be used to estimate the chemical abundances of the ionized gas within the observed galaxy. The majority of the empirical calibrations rely on ratios between bright emission lines, which are defined by the following parameters:
\begin{eqnarray}
R_3&=&  \frac{I([\textsc{O\,iii}]) \lambda 4959+I([\textsc{O\,iii}]) \lambda 5007}{\rm H\beta},\\
R_2 &= & \frac{I([\textsc{O\,ii}]) \lambda\lambda 3727,9}{\rm H\beta},\\
R_{23} & = &  R_3 + R_2, \\
P & = & \frac{R_3}{R_{23}},\\ 
y & = &  \log \frac{R_3}{R_2} = \log \frac{1}{P^{-1}-1},\\
N2 & = & \log{\frac{I([\textsc{N\,ii}]) \lambda 6584}{\rm H\alpha}},\\
O3N2 &= & \log{\frac{I([\textsc{O\,iii}]) \lambda 5007}{I([\textsc{N\,ii}]) \lambda 6584}}, \\
N2O2 & = &\log{\frac{I([\textsc{N\,ii}]) \lambda 6584}{I([\textsc{O\,ii}]) \lambda\lambda 3727,9}}
\end{eqnarray} 
Reviews of the most-common empirical calibrations and their limitations can be found in \citet{2008ApJ...681.1183K}, \citet{2010A&A...517A..85L} and \citet{2012MNRAS.426.2630L}.

Two important issues must be considered when using the SEL technique: 1) To accurately compute some of these parameters (e.g., $R_{23}$, $N2O2$), the emission line fluxes must be corrected for reddening. Hence, this is providing an extra uncertainty of the results; 2) As the intensity of oxygen (or any other heavy element) lines do not monotonically increase with metallicity, some parameters, e.g., $R_{23}$, are actually bi-evaluated. Thus, for the $R_{23}$ index the calibrations must be given for \mbox{\abox$\lesssim$8.0} (low metallicity) and \abox$\gtrsim$8.4 (high metallicity). For these two reasons, it is very convenient --and lately it has been extensively used in the literature-- to resort to parameters that monotonically vary with the oxygen abundance --i.e. parameters that are not bivaluated--, such as $N2$ or $O3N2$, which do not suffer the problems of the reddening correction nor of the bi-evaluation. Nevertheless, some caution has still to be taken using these parameters, as some biases still exist \citep[more details in][and references within]{2012MNRAS.426.2630L}. For example, $N2$ saturates at high metallicities (\abox$\gtrsim$8.65), while $O3N2$ is not valid in the low-metallicity regime (\abox$\lesssim$8.1).

\subsubsection{Adopted oxygen abundances} \label{sec:adopted}

SEL methods based on the Te method consider oxygen abundance data for which the direct measurement of the electron temperature -- derived from line ratios such as \OIII\ $\lambda$4363/$\lambda$5007, see \citet{2006agna.book.....O}-- are available. On the other hand, calibrations based on photoionization methods use photoionization modelling to derive the oxygen abundance directly from emission line ratios.

We here use a combined approach to derive the oxygen abundance of our galaxy sample using several SEL methods. First, we use the empirical calibrations of \citet[hereinafter PP04]{2004MNRAS.348L..59P}, which consider both the $N2$ and $O3N2$ parameters, to get a first estimation of the metallicity. This value is later used to constrain the theoretical \HI\ Balmer lines according to the method explained in the previous section. Once the reddening correction is known, we derived the oxygen abundance using both the \citet{2001A&A...369..594P,2001A&A...374..412P} and \citet[][hereinafter P01 and PT05, respectively]{2005ApJ...631..231P} calibrations --which are based on \Te\ and consider the $R_{23}$ and $P$ parameters -- , and the \citet[hereinafter KK04]{2004ApJ...617..240K} calibrations --which were derived following photoionization models and consider the $R_{23}$ and $y$ parameters. It is well established \citep[e.g.,][]{2010A&A...517A..85L} that the SEL methods based on photoionization models overestimate the oxygen abundances given by the SEL based on the \Te\ method by 0.2--0.4 dex. Hence, we consider the relationship given by \citet{2013ApJ...764..178L} to convert the derived KK04 values into \Te-based values. For the high-metallicity range --\abox$\geq$8.4-- we also use the $N2O2$ --corrected for reddening-- with the calibration provided by \citet[hereinafter KDN2O2]{2002ApJS..142...35K}. We also use the \citet[hereinafter M13]{2013A&A...559A.114M}, based in N2 and in O3N2 parameters and, basically, an update of PP04 calibrations.

\begin{figure}
\centering
\includegraphics[width=1.\linewidth]{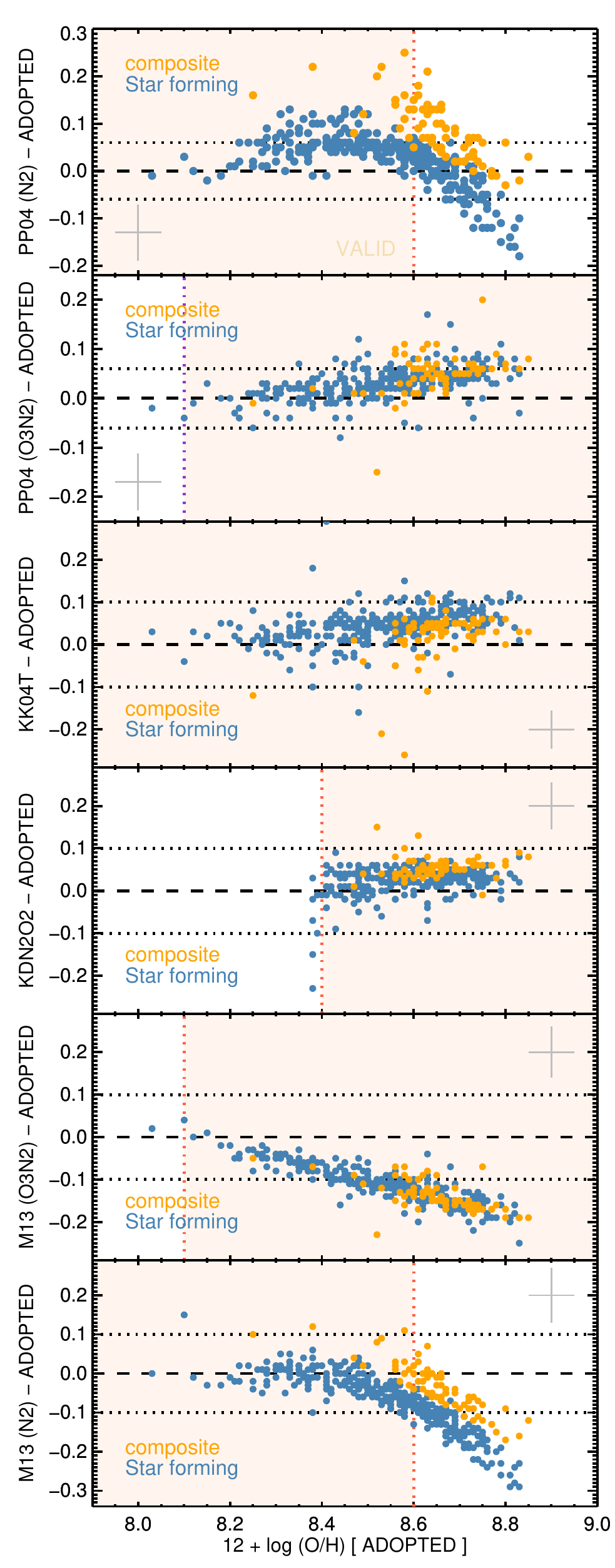}     
\caption{Comparing abundances. 
Blue circles are pure star-forming galaxies, while yellow circles represent galaxies with composite AGN/star-forming nature. Colored background represent the region where the measurement from each calibration is valid.}
\label{fig:figure04}
\end{figure}

\begin{deluxetable}{c@{\hspace{5pt}}c@{\hspace{5pt}}   cc@{\hspace{5pt}} cc@{\hspace{5pt}} cc@{\hspace{5pt}}cc@{\hspace{5pt}}  cc@{\hspace{5pt}}cc@{\hspace{5pt}} cc@{\hspace{5pt}}  cc@{\hspace{5pt}}   } 
\rotate
\tabletypesize{\scriptsize}
\tablecaption{Additional measured lines for the galaxies which we have determined the abundances by the direct method. \label{additional}}
\tablewidth{0pt}
\tablehead{
\colhead{Object} &
\colhead{Nature} &
\colhead{T$_e$ $\rm [O\,III]$ }&
\colhead{T$_e$ $\rm [O\,II]$ } &
\multicolumn{2}{c}{ $\frac {\rm [O\,III]\,\lambda4363}{\rm H\beta}$ } & 
\multicolumn{2}{c}{ $\frac {\rm [O\,III]\,\lambda4959}{\rm H\beta}$} & 
\multicolumn{2}{c}{ $\frac {\rm [N\,II]\,\lambda6548}{\rm H\beta}$} \\
 \noalign{\smallskip}  
\cline{5-6} \cline{7-8} \cline{9-10}  
\noalign{\smallskip} 
&&& & $I$ & $I_0$  & $I$ & $I_0$  & $I$ & $I_0$ \\
}
\startdata
1119   &  C  & 9860     $\pm$ 980    &  9900  $\pm$	680 &  3.72 $\pm$  1.13  & 4.03  $\pm$ 1.23 & 217  $\pm$  13  &  213  $\pm$  14  &   82.4  $\pm$  9.2 &  69.7  $\pm$  7.8 \\
3901   &  SF & 9670    $\pm$ 1410   &  9770  $\pm$	980 &  0.98 $\pm$  0.42  & 1.11  $\pm$ 0.47 & 65  $\pm$  3  &  64    $\pm$  3       &   15.5  $\pm$  2.6 &  11.9  $\pm$  2.0 \\
8719   &  SF & 11050  $\pm$ 870    &  10730 $\pm$	600 &  2.14 $\pm$  0.47  & 2.39  $\pm$ 0.53 & 98  $\pm$  3  &  95    $\pm$  3      &   6.50  $\pm$  0.7 &   5.1   $\pm$  0.6 \\
11172  &  SF & 7940   $\pm$ 1150  &  8560  $\pm$	800 &  0.24 $\pm$  0.12  & 0.26  $\pm$ 0.13 & 31  $\pm$  1  &  31    $\pm$  2      &   50.6  $\pm$  5.6 &  43.8  $\pm$  4.9 \\ 
13072  &  SF & 8970   $\pm$ 710    &  9280  $\pm$	490 &  0.98 $\pm$  0.27  & 1.15  $\pm$ 0.32 & 92  $\pm$  1  &  90    $\pm$  2      &   30.5  $\pm$  1.8 &  21.8  $\pm$  1.3 \\
15132  &  SF & 11350 $\pm$ 1320 &  10940 $\pm$	920 &  2.26 $\pm$  0.70  & 2.56  $\pm$ 0.79 & 91  $\pm$  12  &  89    $\pm$  12   &    1.8   $\pm$  0.3 &   1.4   $\pm$  0.2 \\
\enddata
\end{deluxetable}

\begin{deluxetable}{c@{\hspace{2pt}} c@{\hspace{2pt}} c@{\hspace{2pt}} c@{\hspace{2pt}} c@{\hspace{2pt}} c@{\hspace{2pt}}c@{\hspace{2pt}} c@{\hspace{2pt}} c@{\hspace{2pt}}@{\hspace{2pt}}  c} 
\rotate
\tabletypesize{\scriptsize}
\tablecaption{Direct abundances \label{directoh}}
\tablewidth{0pt}
\tablehead{
\colhead{Object} &
\colhead{Nature} &
\colhead{12 + log(O$^{+}$/H$^{+}$)} &
\colhead{12 + log(O$^{++}$/H$^{+}$)} &
\colhead{12 + log(O/H)} &
\colhead{log(O$^{++}$/O$^{+}$)} &
\colhead{icf (N$^{+}$)} &
\colhead{12 + log(N$^+$/H$^+$)} &
\colhead{12 + log(N/H)} &
\colhead{log(N/O)} \\
}
\startdata
1119  &  C  & 8.01  $\pm$  0.14 & 8.38  $\pm$  0.06  &    8.53  $\pm$  0.06  &     0.36  $\pm$  0.15 &  3.31  $\pm$  0.80  & 7.46  $\pm$  0.08  & 7.98  $\pm$  0.13  & -0.55 $\pm$  0.02 \\
3901  &  SF & 8.24  $\pm$  0.21 & 7.88  $\pm$  0.09  &    8.40  $\pm$  0.15  &    -0.36  $\pm$  0.23 &  1.44  $\pm$  0.23  & 6.95  $\pm$  0.13  & 7.11  $\pm$  0.15  & -1.29  $\pm$  0.03 \\
8719  &  SF & 7.97  $\pm$  0.08 & 7.89  $\pm$  0.03  &    8.23  $\pm$  0.04  &    -0.08  $\pm$  0.08 &  1.83  $\pm$  0.15  & 6.41  $\pm$  0.07  & 6.67  $\pm$  0.08  & -1.56  $\pm$  0.01 \\
11172 &  SF & 8.42  $\pm$  0.22 & 7.90  $\pm$  0.11  &    8.54  $\pm$  0.17  &    -0.52  $\pm$  0.25 &  1.30  $\pm$  0.17  & 7.34  $\pm$  0.13  & 7.45  $\pm$  0.14  & -1.09  $\pm$  0.03 \\
13072 &  SF & 8.14  $\pm$  0.11 & 8.14  $\pm$  0.05  &    8.44  $\pm$  0.06  &    -0.00  $\pm$  0.12 &  1.99  $\pm$  0.28  & 7.18  $\pm$  0.07  & 7.48  $\pm$  0.09  & -0.96  $\pm$  0.01 \\
15132 &  SF & 8.11  $\pm$  0.16 & 7.79  $\pm$  0.06  &    8.28  $\pm$  0.11  &    -0.31  $\pm$  0.17 &  1.48  $\pm$  0.19  & 6.61  $\pm$  0.11  & 6.78  $\pm$  0.12  & -1.50  $\pm$  0.02 \\
\enddata
\end{deluxetable}
\normalsize

For the intermediate metallicity range, $8.0\leq$\abox$\leq8.4$, we use an averaged value between the low- and high-metallicity relationships in all calibrations involving the $R_{23}$ index. This averaged value is actually obtained weighting the equations considering the oxygen abundance provided by the PP04  method. The weighting factor was derived using a linear fit in the range [0,1] between the high-end of the low-metallicity branch --\abox=8.0-- and the low-end of the high-metallicity branch --\abox=8.4--. For example, if an object has \abox=8.3 following the $N2$ and $O3N2$ parameters, we assume that the oxygen abundance derived using equations involving $R_{23}$ is 0.25 of the value of the low-metallicity equation and 0.75 of the value of the high-metallicity  equation. The typical uncertainties of the oxygen abundances obtained following the SEL methods are 0.1\,dex, which is the standard error used for SEL methods (see, i.e., \citealt{2012MNRAS.426.2630L}).

Following the above results, the adopted value for the oxygen abundance in each object is the averaged value of all the valid results, taking also into account that some calibrations are not valid in all the metallicity range (as it was said before, $N2$ is only valid for \abox$\leq$8.65, $O3N2$ can be only applied for \abox$\geq$8.10 and $N2O2$  is only valid for \abox$\geq$8.40). Moreover we have found that P01 and PT05 are not providing good results (see next subsection \ref{te}) and therefore we have not used them for the average calculation.

The results thus obtained are compiled in Table~\ref{ohtable}, where, for each galaxy, named by its SDSS-II SN identification number and the IAU name for Union2.1 SNe~Ia, in column 1, the nature of the spectrum (only SF or C are included in the table) is in column 2, the reddening coefficient c(H$\beta$) is in column 3, and the absorption equivalent width of H$\alpha$ is in column 4. Columns 5 to 12 give the oxygen abundance as \mbox{12+log(O/H)} as obtained from the above described empirical calibrations PP04 using $N2$ (called PP04a), PP04 using $O3N2$ (called PP04b), P01, PT05,  KK04 once corrected for the Te scale (called KK04T), the one from KDN2O2, and those from M13 (M13a from $O3N2$ and M13b from $N2$). Column~13 compiles the finally adopted oxygen abundance.

Once we obtain a value of the oxygen abundance using the SEL methods, and given the \OII/Hb\ and \OIII/\Hb\ ratios and the electron density, we used the five-level program for the analysis of emission-line nebulae included in IRAF NEBULAR task\footnote{We note that we used an updated atomic dataset for O$^+$ and S$^+$ for NEBULAR. The references are indicated in Table~4 of \citep{2004ApJS..153..501G}.} \citep{1995PASP..107..896S} to search for the Te values that best reproduce the observations, also considering the relationship \mbox{Te\OII = 0.7 Te\OIII +3000} \citep{1992AJ....103.1330G}. This exercise also provides an estimation of the ionization parameter, log$({\rm O^{++}/O^{+}})$, in each object. We also derive the nitrogen abundance, N/H, using the \NII/\Hb\ ratio and assuming the standard \citet{1969BOTT....5....3P} ionization corrector factor (icf), N$^+$/O$^+$ = N/O. 

Column 14 in Table~\ref{ohtable} provides the high-ionization electronic temperature, T\OIII. Column 15 gives the log$({\rm O^{++}/O^{+}})$ ratio. Finally columns 16 and 17 provides the nitrogen abundances and the log(${\rm N/O}$) ratio. The N/O ratio does not depend strongly on the electron temperature and hence we tabulate uncertainties for these values.

\begin{figure}
\centering
\includegraphics[trim=0.1cm 0cm 1cm 0cm,clip=True,width=\linewidth]{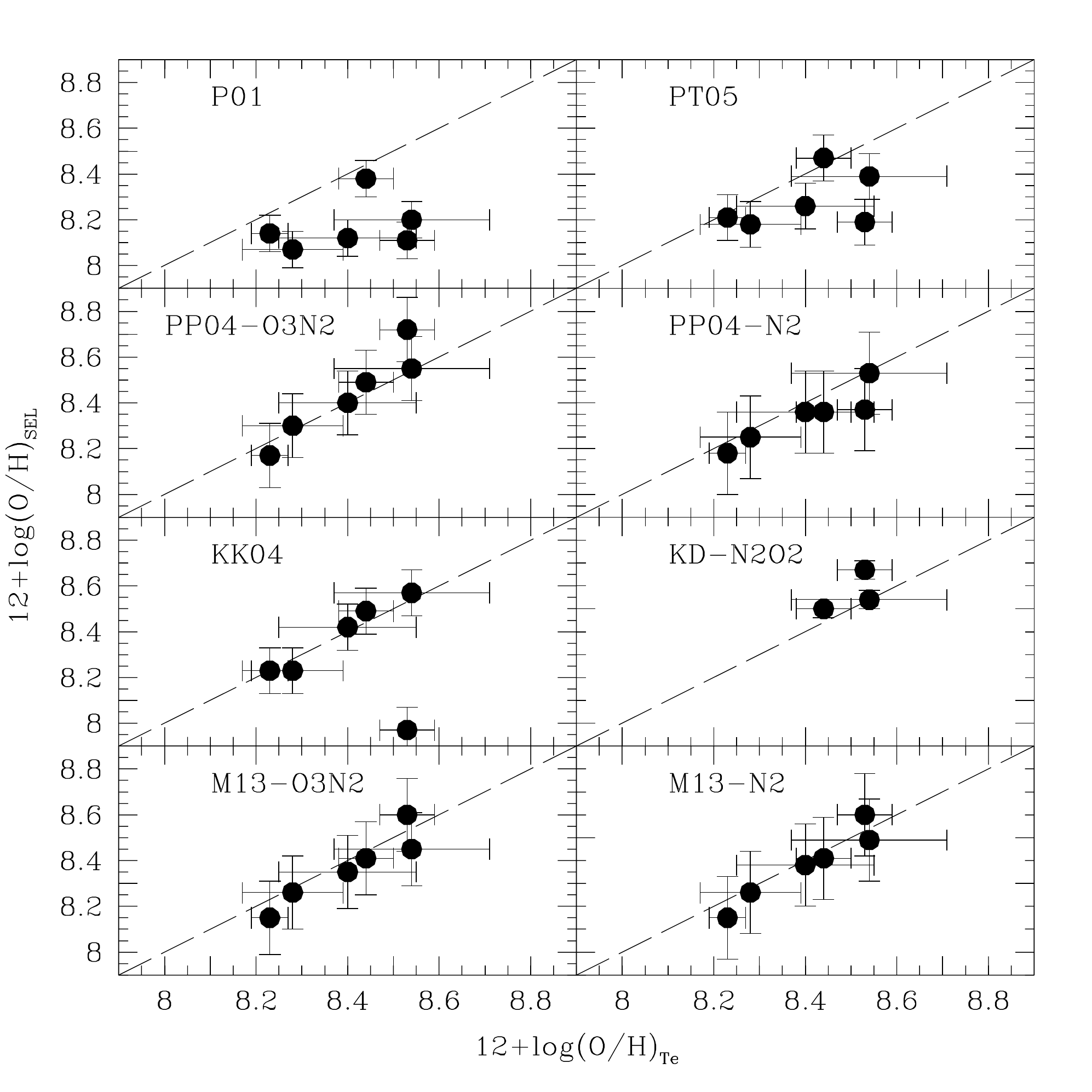}     
\caption{The comparison between the oxygen abundance obtained with the direct method for our six objects where it is possible to
measure \OIII~$\lambda$4363 emission line and 12+log(O/H)$ _{Te}$, with the one obtained with the different described SEL empirical calibrations, 12+log(O/H)$_{SEL}$, as labeled in each panel.}
\label{fig:figure06}
\end{figure}

The errors of the adopted oxygen abundances  consider the dispersion of the values derived following the valid methods, but it will never be inferior to the quadratic sum of the individual errors --e.g., if only 2 independent values are available, the error in the adopted oxygen abundance will not be inferior to $1/\sqrt{1/0.10^2+1/0.10^2}$=0.07\,dex--.

\begin{figure}
\centering
\includegraphics[width=1\linewidth]{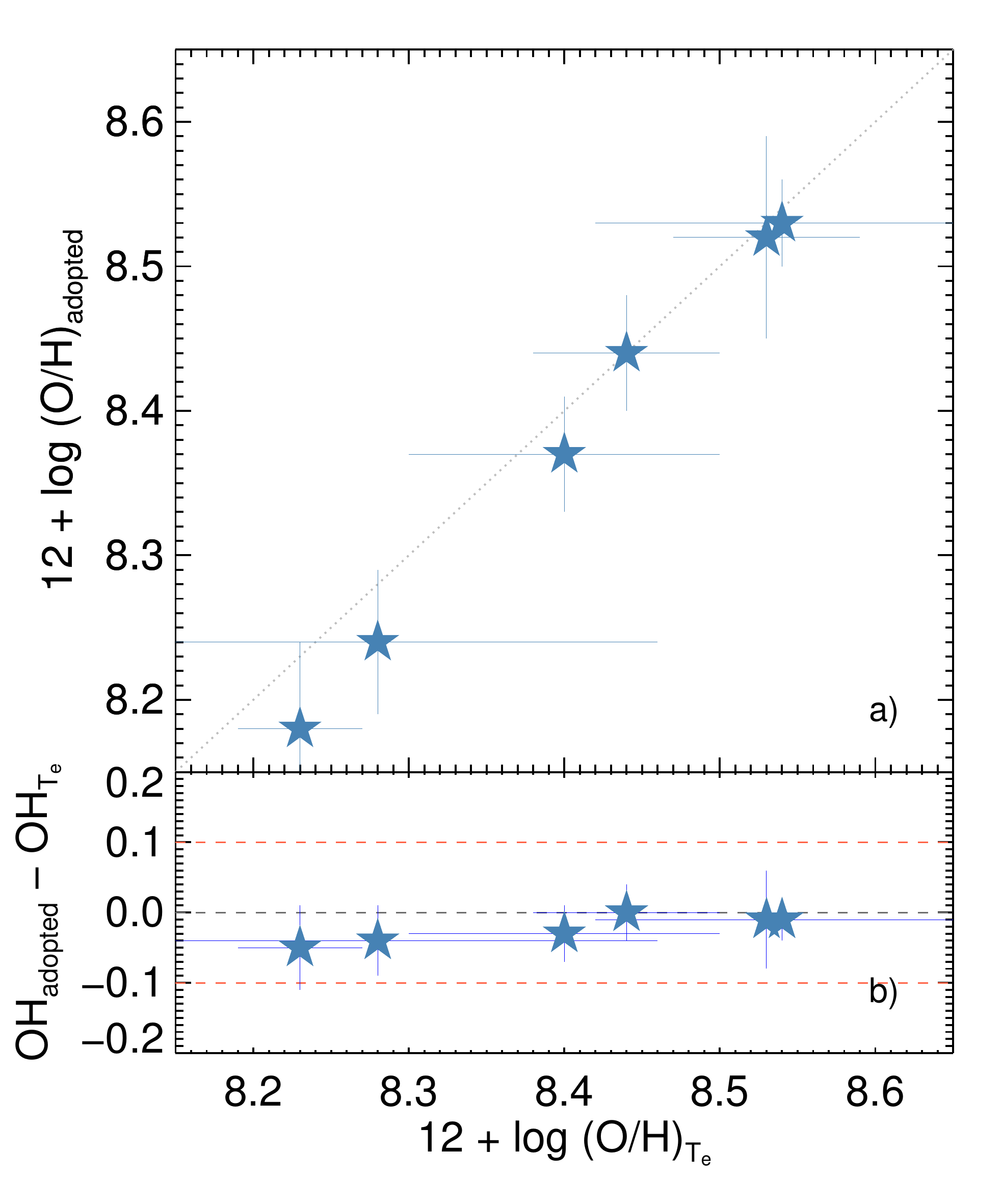}     
\caption{Differences between empirical and direct method abundances for the six galaxies for which this last one is possible. Top panel: the adopted abundance $\rm{12+log(O/H)_{adopted}}$ {\it vs} the direct method one   $\rm{12+log(O/H)_{Te}}$. Bottom panel:
The difference between both values as a function of  $\rm{12+log(O/H)_{Te}}$.}
\label{fig:figure07}
\end{figure}

\begin{figure*}
\centering
\includegraphics[trim=1.5cm 0cm 0.4cm 0.5cm,clip=1,width=\columnwidth]{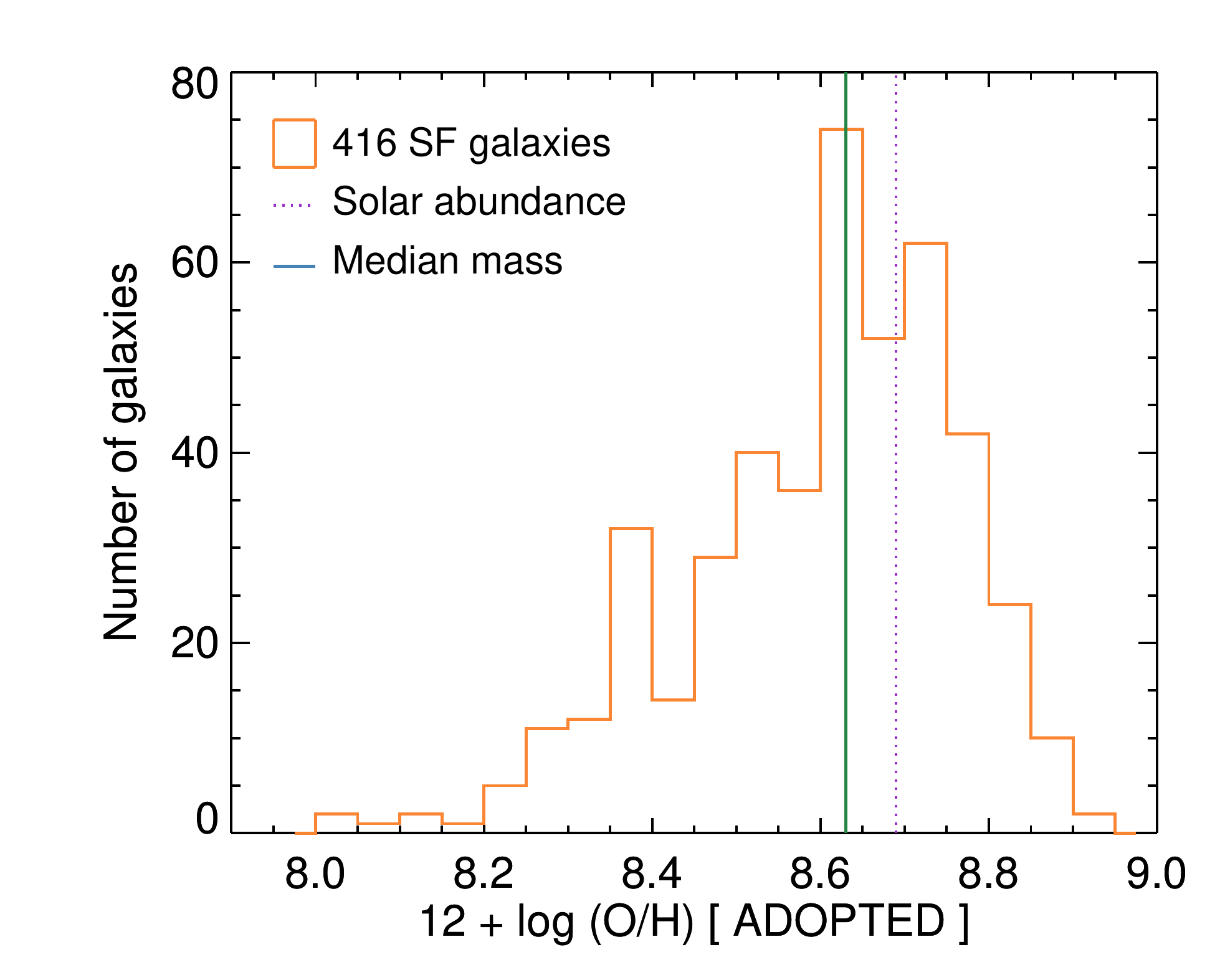}     
\includegraphics[trim=1.5cm 0cm 0.4cm 0.5cm,clip=1,width=\columnwidth]{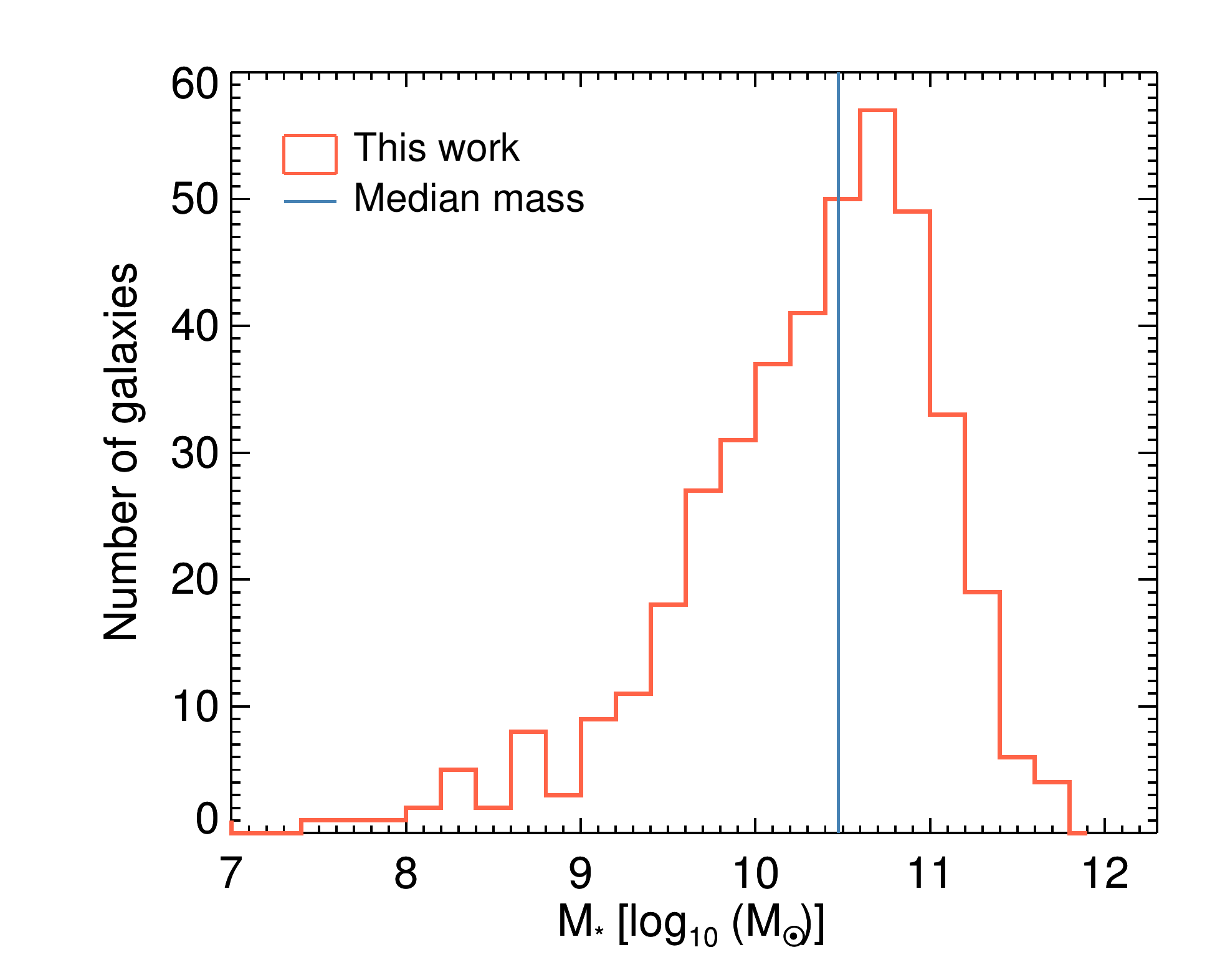}     
\caption{Distribution of determined abundances and stellar masses for the 416 galaxies with host galaxy properties available. In the left panel the solar value given by \citet{2009ARA&A..47..481A} is marked with a long-dashed line.}
\label{fig:figure08}
\end{figure*}

Figure \ref{fig:figure04} resumes the six calibrations we finally use, comparing for the whole sample each calibrated abundance with the adopted one. In the x-axis we have the adopted values while the y-axis represents the differences between each calibration and the adopted values. In each panel the limiting abundances for which the calibration is not longer valid is represented as a dotted line, while the identity or zero value is the dashed line with dotted lines given the 1-sigma values. It is quite evident that using these limits, all the used calibrations are good.

Our final adopted oxygen abundances are shown in the left panel of Fig.~\ref{fig:figure08} as a histogram. The solar value is marked with a long-dashed line. We have a small number of metal-poor galaxies, most of them have around solar metallicity. This bias comes from the SDSS data: small, metal-poor galaxies are not very common to find in their dataset, which is highly dominated by large galaxies, particularly at intermediate- and high-redshifts.

\subsubsection{Comparison to the direct method (Te)}\label{te}

We could detect the faint \OIII~$\lambda$4363 emission line in six galaxies. In those cases, therefore, we derive the oxygen abundance and the N/O ratio using the direct method, i.e., considering the electron temperature of the ionized gas. For this we use again the \citet{1995PASP..107..896S} five-level program included in IRAF NEBULAR task, first to derive Te\OIII\ and then to compute the ionic abundances, O$^+$/H$^+$,  O$^{++}$/H$^+$ and N$^+$/H$^+$ and the total O/H and N/O abundances. The N/O abundances were derived considering the $icf$(N$^+$) assumption given by \citet{1969BOTT....5....3P}.

Table \ref{additional} compiles the additional lines used in these spectra to measure the direct abundances. Column~1 shows the name of the SNe~Ia, column~2 defines the nature of the galaxy (only SF and C are listed in this table). Columns 3 and 4 give electronic temperatures for high and low ionization zones, respectively. The additional emission line intensities as $\frac {\rm [O\,III]\,\lambda4363}{\rm H\beta}$ , $\frac {\rm [O\,III]\,\lambda4959}{\rm H\beta}$ and $\frac {\rm [N\,II]\,\lambda6548}{\rm H\beta}$, in units of $I(H\beta)=100$, with their errors, in columns 5 to 10. For each line, the observed, $I(\lambda$), and theoretical, corrected for reddening, intensities, $I_0$($\lambda$), are shown. Table \ref{directoh} shows the direct abundances determined for these six galaxies.

Figure~\ref{fig:figure06} shows the comparison of the direct method results with those obtained with each one of the calibrations described before. We see that, for these six galaxies, the results provided by the SEL techniques proposed by PP04 (both equations considering the N2 and the O3N2 parameters, respectively), KK04T (using the R23 and $y$ parameters), KDN2O2 (using N2O2 parameter), and M13 (considering the O3N2 and N2 parameters, too), are valid to reproduce the direct method abundances, while the ones from P01 and PT05 are not good enough.

As the two top panels in Fig.~\ref{fig:figure06} show, both P01 and PT05 are providing results that are typically \mbox{0.1 -- 0.2~dex} larger than the average value of the other SEL methods. This was a somewhat surprised result, as previous studies --e.g. \citet{2010A&A...517A..85L}-- reported that P01 and PT05 were actually providing better results than the other SEL methods when compared to their Te-based oxygen abundances. We suspect that the reason of this behavior is consequence of our SDSS galaxy sample being very different to the galaxy sample whose \HII\ regions were used to obtain the P01 and PT05 calibrations, introducing an additional important bias when deriving their oxygen abundances \citep[see][for an extended discussion]{2010IAUS..262...93S}.

Therefore, we adopt for the oxygen abundance of each galaxy the averaged value of all the valid results, taking also into account that some SEL calibrations can not be used in all the metallicity range, since, as we said before, N2 is only valid for \abox$\leq 8.65$, O3N2 can be only applied for \abox$>8.10$ and N2O2 is only valid for \abox$>8.40$.

Figure \ref{fig:figure07} shows the differences between the final adopted abundances and those from the direct method. It is clear that our method is good enough to obtain abundances as the ones estimated with the electronic temperature from the \OIII~$\lambda$4363  emission line. Therefore, we adopt this same criterion to  estimate the oxygen abundances for the whole sample.

\begin{figure*}
\centering
\includegraphics[width=1\linewidth]{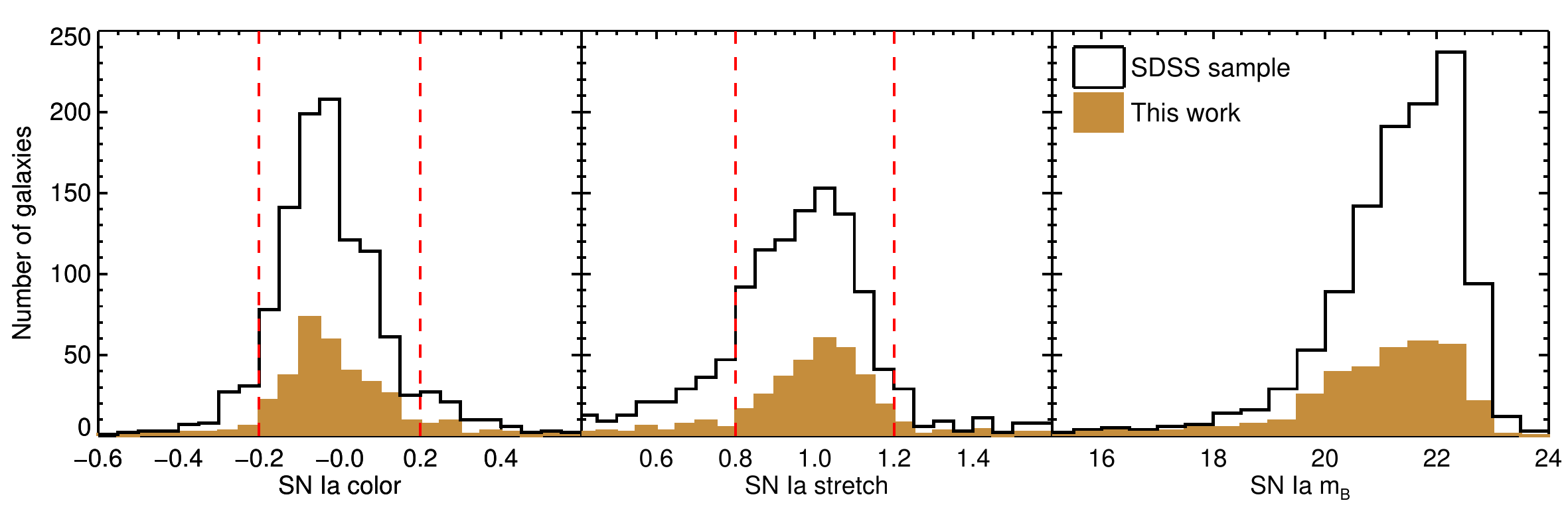}
\caption{Properties of the SNe~Ia SDSS sample and our host galaxy sample (golden shaded histograms). Left: Distribution of SNe~Ia color, $c$. Center: SNe~Ia stretch, $s$, distribution. Right: Distribution of apparent magnitudes, $m_{B}$. Red dashed lines define limits for color and stretch.}
\label{fig:figure09}
\end{figure*}

\begin{table}
\caption{Number of SNe/galaxies remaining after each step.}
\label{tab:sample}
\begin{center}
\begin{tabular}{lccc}
\hline\hline
Step & SDSS& Union2.1 & Total\\
 \hline
Initial            & 1466 & 580 & 2046 \\
Spectra in DR12    & 1128 &  60 & 1188 \\
Emission line cuts &  424 &  23 &  447 \\
SF ionization      &  397 &  19 &  416 \\
LC quality cuts    &  327 &  19 &  346 \\
stretch/color cuts &  245 &  18 &  263 \\
\hline
\end{tabular}
\end{center}
\end{table}

\subsection{Stellar masses}

\begin{table*}
\caption{SNe~Ia LC parameters \label{tab:LC_parameters}}
\begin{center}
\begin{tabular}{cccccccc}
\hline\hline
 {Object} & $z$ & {$s$} & {$c$} & {$m_{B}$} & {${\rm log}[M_{*}/M_{\odot}]$} & {SNR}  \\
\hline	
5	&	0.14195 &	1.03	$\pm$	0.02	&	-0.08	$\pm$	0.02	&	20.11	$\pm$	0.03	&	10.714	&	5.7		\\
10	&	0.05878	&	1.12	$\pm$	0.04	&	0.76	$\pm$	0.02	&	20.16	$\pm$	0.03	&	8.408	&	3.9		\\
30	&	0.14276	&	1.09	$\pm$	0.02	&	-0.12	$\pm$	0.01	&	19.51	$\pm$	0.02	&	10.976	&	6.2	\\
83	&	0.05025 &	0.98	$\pm$	0.02	&	-0.1	$\pm$	0.01	&	17.47	$\pm$	0.02	&	10.132	&	5.5	\\
128	&	0.15179	&	1.00	$\pm$	0.02	&	-0.09	$\pm$	0.02	&	20.04	$\pm$	0.02	&	8.389	&	2.8	\\
...\\
\hline
\end{tabular}
\end{center}
\end{table*}

\begin{figure}
\centering
\includegraphics[trim=0.1cm 0cm 0cm 0cm, clip=True,width=1\linewidth]{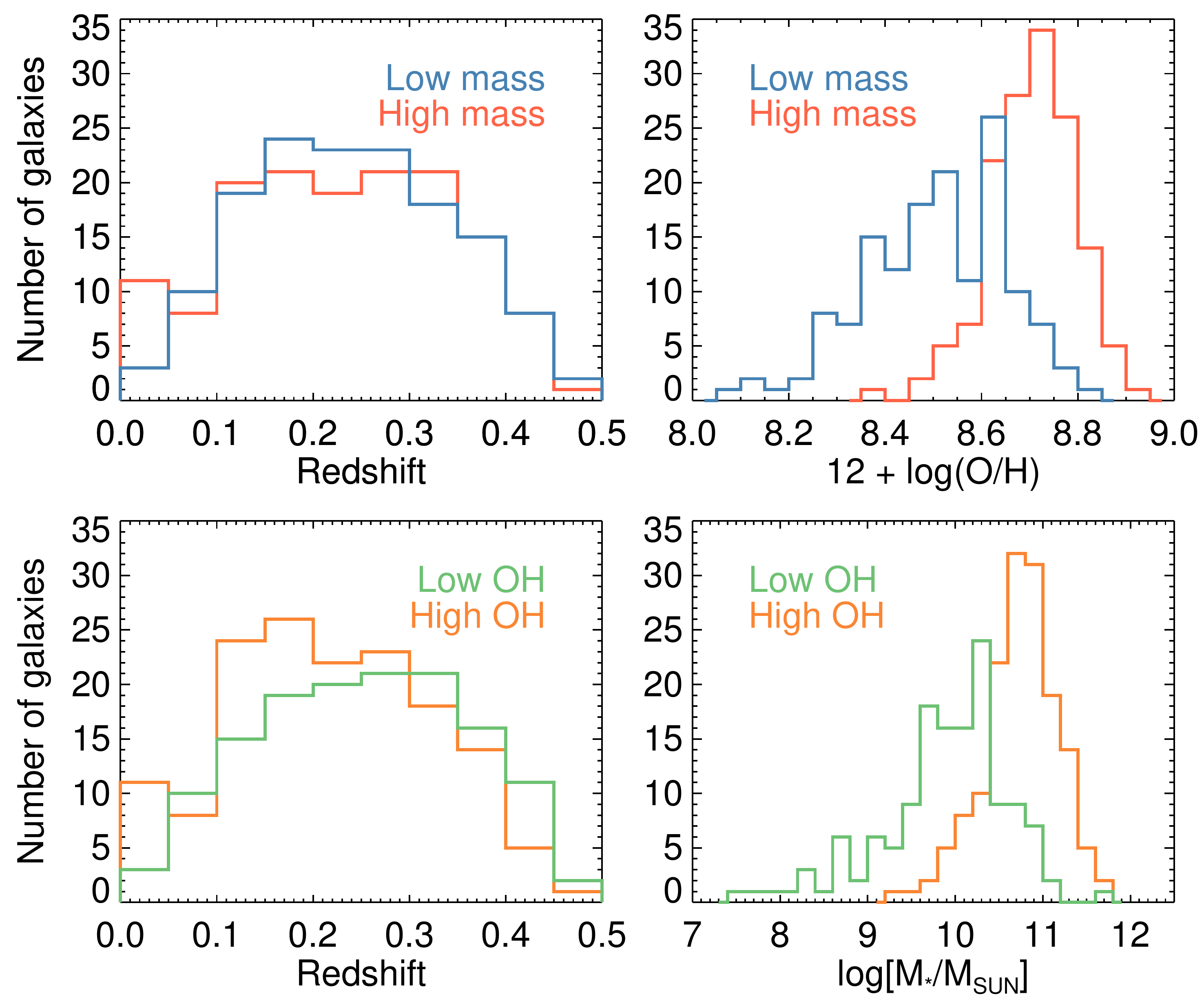}  
\caption{
Distribution of our final sample of host galaxies. Left: distribution in redshift divided in two bins following  the low and high stellar mass (top), or following the low and high oxygen abundances (bottom). Right: distribution in oxygen abundances with two stellar mass bins (top) and in stellar mass with two bins in metallicity (bottom), as labeled.}
\label{fig:figure10}
\end{figure}

\begin{figure*}
\centering
\includegraphics[width=1\linewidth]{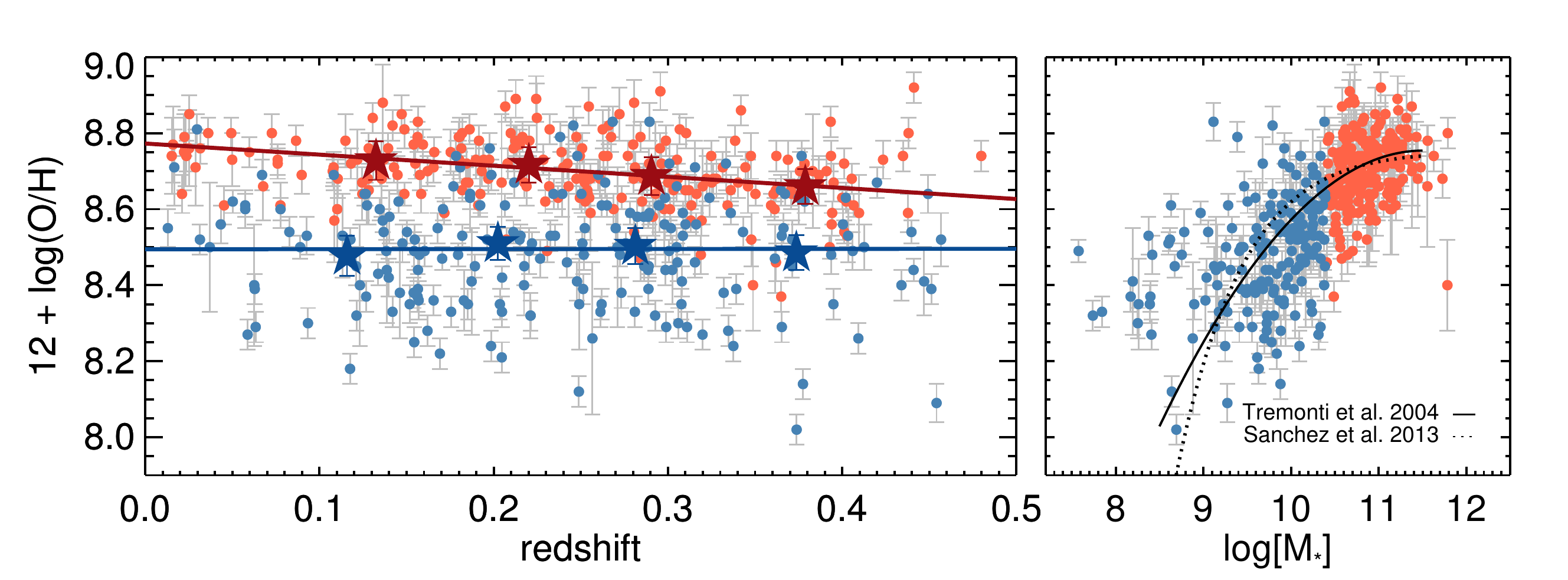}     
\caption{ 
Left. The oxygen abundance as a function of the redshift. Right. The mass-metallicity relation. Red and blue points represent galaxies with stellar mass higher and lower than $\log{M_{*}}=10.4$. Stars (red and blue) represent the abundances binned in four groups of similar sizes for each stellar mass set. The solid and dotted lines in the right panel are the fits found by \citet{2004ApJ...613..898T} and \citet{2013A&A...554A..58S}, respectively.}
\label{fig:figure11}
\end{figure*}

\begin{figure}
\centering
\includegraphics[trim=0.6cm 0cm 0.5cm 0cm,clip=True,width=1\linewidth]{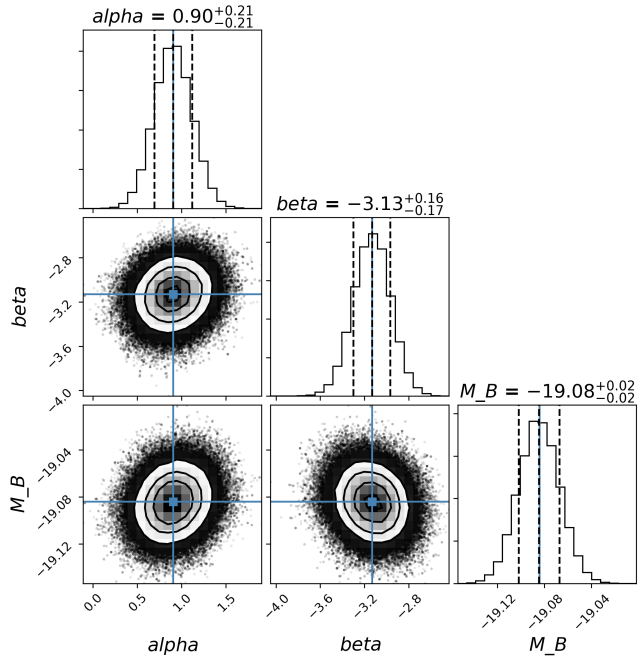}     
\caption{Corner plot of the posterior distributions for the three free parameters in our EMCEE fit ($\alpha$, $\beta$, M$_0$).}
\label{fig:figure12}
\end{figure}

\begin{figure}
\centering
\includegraphics[width=\linewidth]{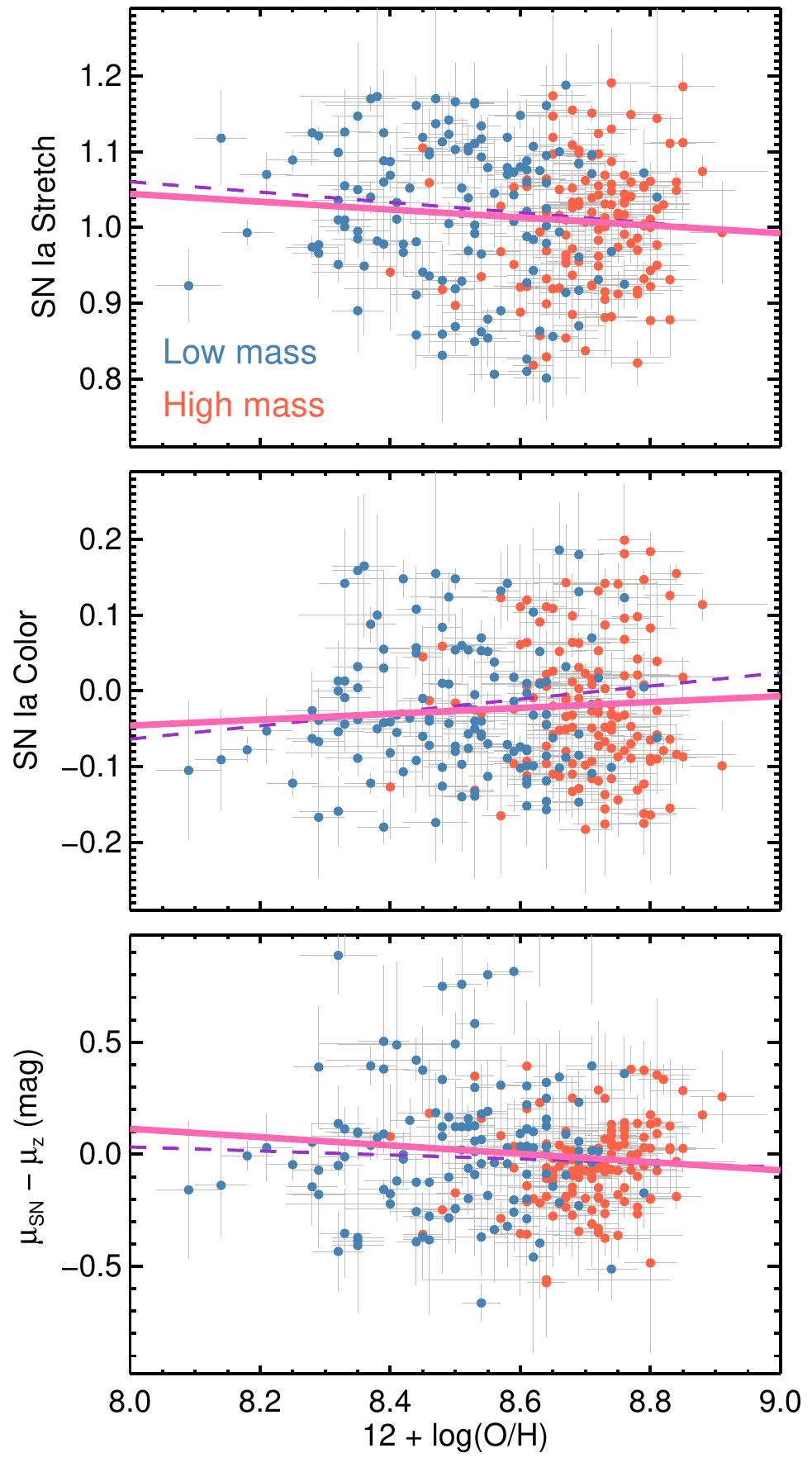}     
\caption{Dependences between stretch (top panel), color (middle panel), and Hubble residuals (bottom panel) oxygen abundance ($12 +log(O/H)$ R) for our  sample of SNe~Ia .}
\label{fig:figure14}
\end{figure}

We have estimated the stellar masses of our galaxies from their  spectral energy distribution. For that we used {\sc STARLIGHT} \citep{2005MNRAS.358..363C,2006MNRAS.370..721M,2007MNRAS.381..263A,2009RMxAC..35..127C}, a program that fits a rest-frame galaxy spectrum (O$_{\lambda}$) in terms of a model (M$_{\lambda}$) built by a nonparametric linear combination of single stellar-population (SSP) spectra from a base spanning different ages and metallicities. The contribution of the different SSPs that best describe the original spectrum can be used to study the properties of the galaxy stellar populations and to estimate stellar velocity fields. In our case we have used the selection of the SSP model bases described in \cite{2012A&A...545A..58S} and already used in other works (e.g.  \citealt{2014A&A...572A..38G,2016MNRAS.455.4087G}), which consists of 66 components with 17 different ages (from 1 Myr to 18 Gyr) and four metallicities ($Z=$0.2, 0.4, 1.0 and 2.5 Z$_{\sun}$, where Z$_{\sun}=0.02$) coming from a slightly modified version of the models of \cite{2003MNRAS.344.1000B}, replacing STELIB by the MILES spectral library \citep{2006MNRAS.371..703S}, Padova 1994 evolutionary tracks, a \cite{2003PASP..115..763C} initial mass function (IMF) with lower and upper mass limits at 0.1 and 100\,M$_{\sun}$, respectively, and new calculations of the TP-AGB evolutionary phase for stars of different mass and metallicity by \cite{2007A&A...469..239M} and \cite{2008A&A...482..883M}. Dust effects are modeled as a foreground screen with a \citet{1989ApJ...345..245C} reddening law with $R_{V} = 3.1$.  

The output from STARLIGHT is the best contribution of SSPs of different ages and metallicities that reproduces the input spectrum. The total stellar mass is recovered by combining the mass-to-light ratio of the different SSPs contributing to the best fit. This stellar mass refers to the integrated star formation history and, therefore, includes stellar masses of different ages. A detailed analysis of the star formation histories of these galaxies will be presented elsewhere (Galbany et al. in prep.). The right panel of Figure \ref{fig:figure08}, shows the distribution of these stellar masses for our sample of galaxies. 


\section{SN LC parameters} \label{sns}

We fitted all SN Ia LCs with {\sc SiFTO} \citep{2008ApJ...681..482C} in order to obtain the apparent magnitude at peak (already corrected for Milky Way extinction), the width of the LC or stretch, $s$, and the SNe~Ia color at maximum, $c$. SiFTO has been successfully used in several studies \citep{2010MNRAS.406..782S,2011ApJS..192....1C}. We discarded SNe Ia that did not pass the following criteria in their LCs:
1) at least 4 points between -10 and 35 days from peak;
2) at least 2 points between -10 and 5 days from peak; 
3) at least 1 point between +5 and +20 days from peak; and
4) at least 1 point between -8 and +8 days from the peak in two different filters.
From our 416 SNe~Ia hosted in SF or C galaxies, 83\% pass these quality cuts, the sample being now reduced to 346 objects. To the previous described quality cuts we have also applied cuts on color (-0.2$<c<$0.2) and stretch (0.8$<s<$1.2) parameters to keep a sample of SNe Ia as much similar as possible to a {\it normal} SN Ia. After this consideration, the number of remaining SNe~Ia is 263.

Similarly to Fig.~\ref{fig:figure08}, Fig.~\ref{fig:figure09} present the distributions of the SNe~Ia LC parameters, color $c$, stretch $s$ and apparent magnitude $m_{B}$ and shows again no selection biases in our sample. We verify with this figure that our sample is not biased compared with the total SDSS sample.

Table~\ref{tab:sample} summarizes the number of SNe/galaxies useful for our analysis after each cut, and Table~\ref{tab:LC_parameters} lists redshift, light-curve parameters and host galaxy stellar mass for the remaining SN~Ia in our final sample.


\section{Results and Discussion} \label{Section5}

\subsection{Galaxy mass and metallicity distributions}

We divided the sample into two groups according their mass and metallicity:  galaxies are divided in two bins of low and high mass at  $\log{M_{*}/M_{\sun}}=10.4$. This limiting value is the median mass of our galaxy sample, in order to have the same size in both groups. Similarly the metallicity division is done at $12+\rm{log(O/H)}=8.64$ which is the median value of the oxygen abundance distributions (and almost the Solar value).

In Fig.\ref{fig:figure10} we show the distributions in redshift, mass and metallicity of our subsamples as described above. Left panels show the redshift distributions divided by mass (top panel) and metallicity (bottom panel), which demonstrate that there is no bias on metallicity in our sample with redshift. 

Right panels show the mass and metallicity distributions divided into two bins according to the other corresponding parameter of metallicity and mass, respectively. Both of them show the effect of the mass-metallicity relation with more massive galaxies in the high metallicity extreme and no metal-rich galaxies below $\log{(M_{*})}=9.2$.

Now that we have all quantities well determined for SNe~Ia and for their host galaxies, we are going to analyze the dependences among them, mainly to see if a correlation with the oxygen abundances appears.

\begin{table*}
\caption{Least squares straight line fitting parameters: $y=a\,x +b$ for results of Fig.~\ref{fig:figure14}  }
\label{tab:fits}
\begin{center}
\begin{tabular}{llcccc}
\hline\hline
Host parameter& SN parameter& intercept & slope & Significance & \% slopes\\
\hline
POLYFIT   &s & 1.605 $\pm$  0.080 & -0.068 $\pm$  0.009 &  7.35$\sigma$ & $-$ \\
          &c &-0.765 $\pm$  0.084 &  0.088 $\pm$  0.010 &  8.96$\sigma$ & $-$ \\
          &HR& 0.751 $\pm$  0.347 & -0.090 $\pm$  0.040 &  2.23$\sigma$ & $-$ \\
\hline
LINMIX    & s &  1.461 $\pm$ 0.326 & -0.052 $\pm$  0.038 &  1.38$\sigma$ & 91.6\\
          & c & -0.357 $\pm$ 0.340 &  0.039 $\pm$  0.039 &  0.99$\sigma$ & 84.3\\
          &HR &  1.600 $\pm$ 1.055 & -0.186 $\pm$  0.123 &  1.52$\sigma$ & 93.6\\
\end{tabular}
\end{center}
\end{table*}
\begin{figure*}
\centering
\includegraphics[width=\linewidth]{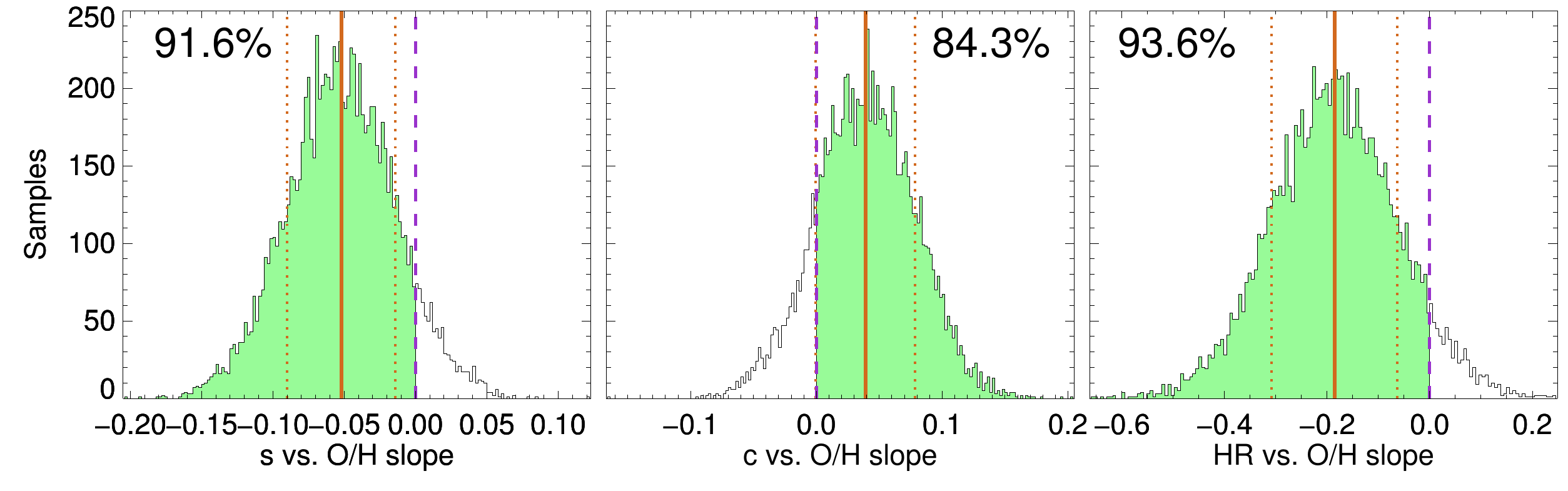}     
\caption{LINMIX posteriors for the slopes of the three fits shown in Figure \ref{fig:figure14}. The percents correspond to the samples consistent with slopes different from zero and in the same direction of the average slope. All distributions are consistent with a Gaussian distribution, which is confirmed by the practically null difference between the mean and the median of each distribution.}
\label{fig:linmix}
\end{figure*}

\subsection{Dependence of metallicity with redshift} \label{sec:oh-z}

In Fig.~\ref{fig:figure11}, left panel, we show the oxygen abundances we obtain with the method described in section \ref{sec:adopted} as a function of redshift. We perform least-squares straight lines for both bins, finding the following equations:
\begin{equation}
\small
12 + \log{\rm{(O/H)}_{hi-mass}} = -0.29 (\pm 0.26)\times z + 8.77(\pm0.07),
\end{equation}
\begin{equation}
\small
12 + \log{\rm{(O/H)}_{lo-mass}} =  0.002 (\pm 0.253)\times z + 8.49(\pm0.06).
 \end{equation}

Low mass galaxies seem to show no variation of metallicity with redshift, but it is necessary to take into account that the dispersion is larger than in the massive galaxy group. Besides that, there is a possible selection effect that would increase the number of massive/bright galaxies observed at higher redshift (Malmqvist bias effect). However, it is also  true that these low stellar mass galaxies have usually stronger star formation rate (at least in this redshift range) and, therefore, the emission lines will be more intense, thus reducing, at least partially, the possible bias. On the other hand, we should emphasize that this figure does not have to be interpreted as an evolutionary scheme. Each galaxy has its own elemental abundance, but the evolutionary track of each galaxy may be completely different than the global trend shown by the whole sample. This finding would suggest that SNe in low-mass galaxies are possibly better standard candles, in line with findings by \cite{2013A&A...560A..66R}.

In the right panel we show the classical mass-metallicity relation found for our sample, and compare it to the relation found by \citet{2004ApJ...613..898T} using the full SDSS sample (scaled to the \Te\ abundance scale) and by \citet{2013A&A...554A..58S}, for the local Universe (0.005 $<z<$ 0.03). Our results are in agreement with both correlations, which is expected in the first case because we use spectra from the same SDSS sample, but also in the second case even though the redshift range of our galaxies is higher.

\subsection{Hubble residual dependence on O/H} \label{HR}

We construct a Hubble diagram from our SN light-curve measurements assuming a fixed standard flat $\Lambda$CDM cosmology with $H_{0}=70\,km\,s^{-1}\,Mpc^{-1}$ and $\Omega_{M}=0.295$ \citep{2014A&A...568A..22B}. To this end, as a first step we minimized the likelihood function,
\begin{equation}
ln(\mathcal{L})=-\frac{1}{2} \sum_{SN} \left\lbrace \frac{[m_i^{obs} - m_i^{model}]^2}{\sigma_{tot}^2} + ln(\sigma_{tot}^2) \right\rbrace,
\end{equation}
where m$_i^{obs}$ is the apparent peak magnitude from SiFTO, and m$_i^{model}$ is,
\begin{equation}
m^{model} = \mu^{\Lambda CDM}(z) + M_B - \alpha  (s - 1) + \beta  c,
\end{equation}
and the error budget contains the uncertainties of each term,
\begin{equation}
\sigma_{tot}^2= \sigma_{m_i}^2 + (\alpha \sigma_s)^2 + (\beta \sigma_c)^2 + (\sigma_{sys})^2.
\end{equation}
Due to our imperfect knowledge of SN Ia physics, commonly a $\sigma_{sys}$ free parameter is included in the error budget to allow for further unaccounted magnitude variations. Given the nature of our minimization (we are fixing the cosmological parameters), $\sigma_{sys}$ does not only account for intrinsic variations but for the fact that the assumed cosmological parameters have no errors.

Following \citet{2017MNRAS.472.4233D}, the best fit parameters from the minimization of the likelihood function are then used as a first guess to initialize a Bayesian Monte Carlo Markov Chain (MCMC) simulation using the Python-based package EMCEE \citep{2013PASP..125..306F}. EMCEE uses an ensemble of walkers which can be moved in parallel to explore probability space, instead of a single iterative random walker (Goodman-Weare algorithm versus Metropolis-Hastings algorithm). Each walker attempts a specified number of steps, first proposing a new position, and then calculating the likelihood of the new position based on the data. The walker has a chance to move to the new position based on the ratio of the likelihood of its current position to that of the new position. After a specified number of steps, the algorithm is completed and the position and likelihood of each walker as a function of step number is written out. These positions and likelihoods are used to generate the posterior probability distribution function for each model parameter. In our fit, we used 500 walkers and 1000 steps, although discarding the initial 100 steps to flatten the list of samples. Our prior probability distribution is defined to have flat probability for 0.0$<\alpha<$3.0 and 1.0$<\beta<$6.0 and -21.0$<M_B<$-17.0 but otherwise have zero probability. We also attempted to use Gaussian priors, however no differences were found in the fit parameters. Following this procedure, in Figure~\ref{fig:figure12} we show a corner plot with the parameter distributions from EMCEE, which gives the most probable values for $M_{B}$, $\alpha$ and $\beta$. We obtain the following mean values and standard deviations from the posterior distributions,
\begin{equation}
(M_{B}, \alpha, \beta)=(19.085 \pm 0.017, 0.905 \pm 0.213, 3.131 \pm 0.169).
\end{equation}

Fig.~\ref{fig:figure14} shows stretch, color, and HR, as a function of oxygen abundance, $12+log(O/H)$. All points are colored by the mass group they belong, which is divided by the median host galaxy stellar mass. To study possible correlations we use the {\sc LINMIX} code, which is more robust than a simple linear fit because it takes into account the errors in both, $x$ and $y$ variables. This method computes a probability function for the data set, using a Markov Chain Monte Carlo (MCMC) algorithm. In each panel we show the best fit from {\sc LINMIX} (thick pink solid line) together with a simple polynomial fit with the {\sc polyfit} IDL library (thin dashed purple line). In Table \ref{tab:fits} we give our results for the fitting parameters of the dependence of $s$, $c$ and $HR$ on oxygen abundance for both methods, and in Figure \ref{fig:linmix} we show the posterior distributions for the slopes given by {\sc LINMIX}. Given that all distributions are consistent with Gaussians, we consider the significance of our relations statistically robust.

\begin{figure}
\centering
\includegraphics[trim=2.2cm 0cm 0cm 0cm,clip=true,width=\columnwidth]{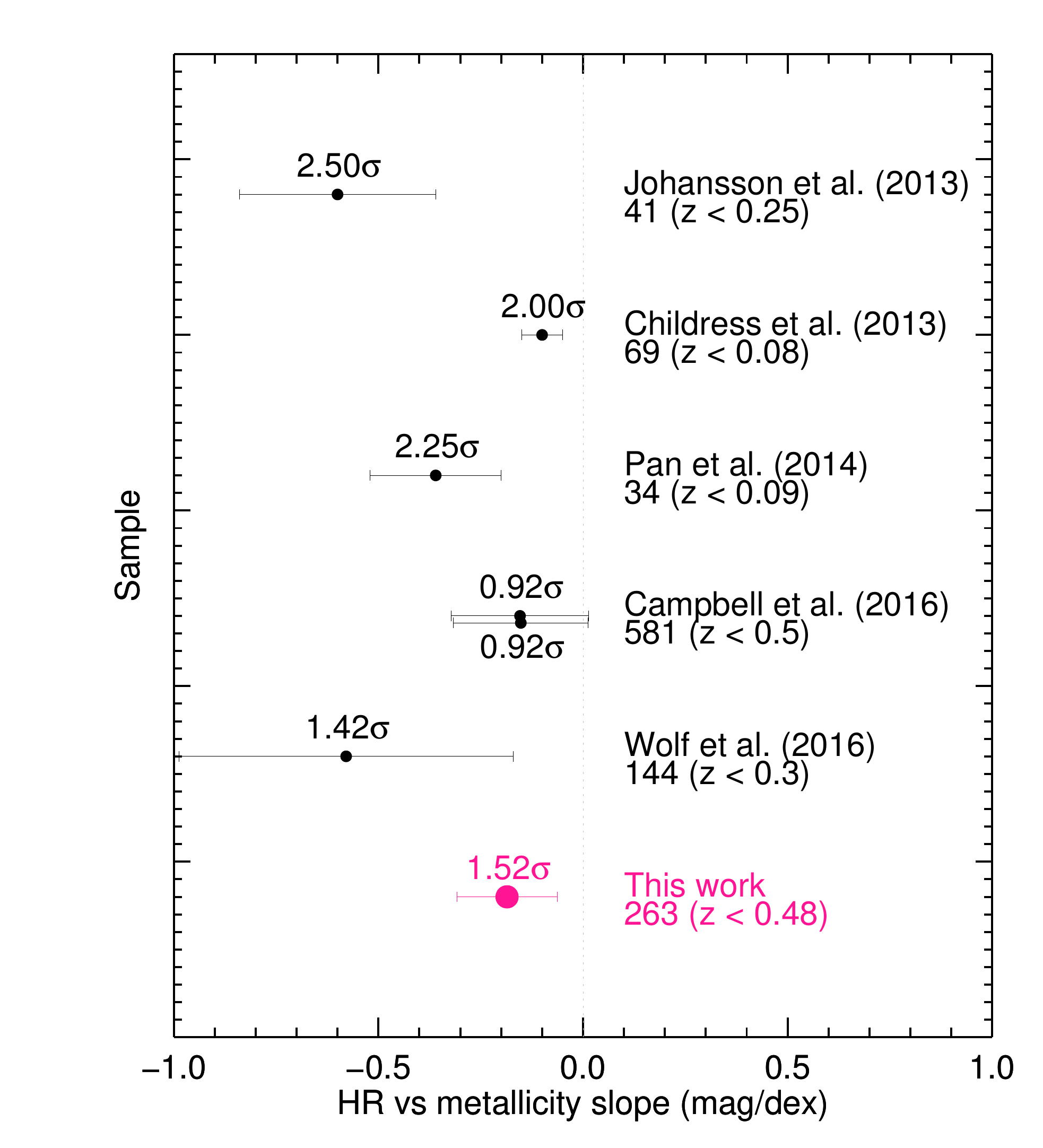}     
\caption{Comparison between our slope of the relationship HR {\it vs} the host galaxy oxygen abundance in mag\,dex$^{-1}$ and others from the literature. }
\label{fig:figure15}
\end{figure}

We find that the SNe~Ia stretch $s$ and color $c$ show clear correlations with metallicity, independently of the method to fit the data, and in agreement with results from \citet{2013ApJ...770..108C} and \citet{2014MNRAS.438.1391P}: metal-poor galaxies host higher stretch (narrower LCs) and with more negative colors (bluer) SNe~Ia. On the other hand, the most interesting result is the correlation between the residuals of the HD and the oxygen abundance, with a significance of 1.52$\sigma$ and a slope of -0.186$\pm$0.123 mag\,dex$^{-1}$. In this sense, most metal-rich galaxies host slightly brighter SNe~Ia {\it after standardization}, as other authors already reported \citep{2013MNRAS.435.1680J,2013ApJ...770..108C,2014MNRAS.438.1391P,2016MNRAS.457.3470C,2016ApJ...821..115W}. This trend is in line with the difference in the Hubble residuals (step) between the two halves of the sample divided at the median value of the metallicity at 8.64 dex. We find that the metal-poor half of the sample has a positive HR of 0.080, while the metal-rich half has a negative HR=-0.026 (step of 0.106 mag). We compare our results with recent works from the literature in Fig.~\ref{fig:figure15}. 

The slope obtained for HR {\sl vs} $12+\log\rm{(O/H)}$, $-$0.186\,mag\,dex$^{-1}$, implies that SNe Ia located in metal-rich environments have smaller distance modulii than what it is expected for their redshift. In Paper I and II, we determined SN Ia absolute magnitudes by calculating distances to their host galaxies using independent estimators (e,g. Cepheids \& Tully-Fisher). We found a difference in this independent M$_B$ measurement of 0.14 $\pm$ 0.10 mag between the metal-rich and metal-poor SNe Ia bins, which is in total agreement with the results found here. Moreover, this difference is also of the order of theoretical explosion models  ($\sim$0.10-0.20 mag) from \cite{2010ApJ...711L..66B}. However, all cosmological analyses, including this presented here, are still assuming a fixed M$_B$ for all SNe Ia.

\begin{figure}
\centering
\includegraphics[trim=0.4cm 0cm 0.6cm 0cm,clip=True,width=1\linewidth]{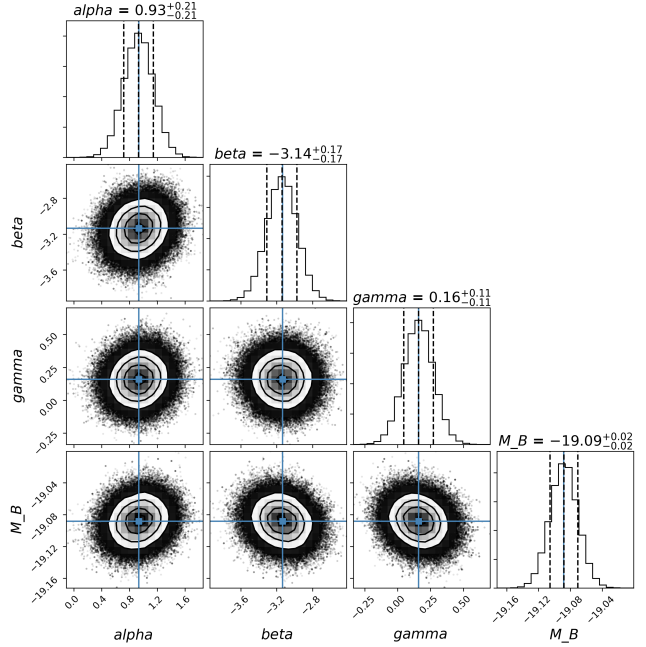}     
\caption{Similarly to Figure \ref{fig:figure12}, corner plot of the EMCEE posterior distributions for our new model including the $\gamma$ parameter which corrects for metallicity differences.}
\label{fig:newcorner}
\end{figure}

The way to solve this problem and obtain flat HR resides in adding a term in the standardization that includes this M$_B$ dependence on metallicity. To this end, we repeated the EMCEE minimization but now adding a third parameter in our standardization model,
\begin{equation}
m^{model} = \mu^{\Lambda CDM}(z) + M_B - \alpha  (s - 1) + \beta  c - \gamma (OH - 8.64),
\end{equation}
where 8.64 dex corresponds to the median oxygen abundance in our sample, and including the corresponding term in the total uncertainty budget,
\begin{equation}
\sigma_{tot}^2= \sigma_{m_i}^2 + (\alpha \sigma_s)^2 + (\beta \sigma_c)^2 + (\gamma \sigma_{OH})^2 + (\sigma_{sys})^2.
\end{equation}
The new best fit parameters determined from the posterior distributions are now,
\begin{multline}
(M_{B}, \alpha, \beta, \gamma) = (19.088 \pm 0.017, 0.932 \pm 0.214, \\ 3.140 \pm 0.170, 0.160 \pm 0.113).
\end{multline}
The change in M$_{B}$, $\alpha$, and $\beta$ is minimal, and we obtained a $\gamma$ value compatible with being non-zero. We also confirmed that using this new model the dependence with metallicity vanishes (slope posterior centered at zero). 

\section{Conclusions and future work}

Our project is divided in three parts: low, intermediate, and high redshift SNe Ia. After the analysis of a sample of SNe Ia in the local Universe (Paper I and II), here we presented the analysis performed to the intermediate redshift sample. Our conclusions are listed as follows:
\begin{enumerate}
\item We have estimated the oxygen abundances by carefully measuring the emission lines of a large sample of SN Ia host galaxies, for which the LC parameters and apparent magnitudes are available.
\item  We constructed a HD and fit the usual parameters $M_{B}$, $\alpha$ and $\beta$ necessary to calculate the distance modulus for each SN Ia of our sample as: $\mu-(M_{B}+\alpha (s-1) -\beta C$). Our values are: $M_{B}=-19.085(\pm 0.017)$, $\alpha=0.905 (\pm 0.213)$  and $\beta=3.131(\pm 0.169)$, in good agreement with results from other authors.
\item The Hubble residual to a fixed $\Lambda$CDM cosmology shows a 1.52$\sigma$ correlation with the global host galaxy oxygen abundance. The slope and significance of this correlation is similar to others given in the literature.
\item The negative slope in the HR correlation with oxygen abundance is explained as fainter SNe Ia explode in metal-richer environments, in the same way as we found in Paper I and II for the local Universe. 
\item We proposed the addition of a third parameter in the SN Ia standardization ($\gamma$=0.160$\pm$0.113), that accounts for the metallicity and corrects for the observed trend. 
\end{enumerate}

The main caveat of the addition of a third parameter in the standardization that accounts for the metallicity is that it will have a dependence on $z$, as we demonstrated in section \ref{sec:oh-z}, due to the chemical enrichment evolution of the Universe. In addition, and in contrast with our work at low-$z$ in Paper I and II, we are here using metallicities measured from a fiber spectrograph that covers different fractions of the galaxy depending on its size and redshift. On average, galaxies have negative metallicity gradients, which translates in central regions having higher metallicities than outer regions \citep{2014A&A...563A..49S}. For this reason, fiber spectroscopy is known to suffer from aperture effects that have to be corrected for in order to obtain the {\it characteristic} metallicity of the galaxy \citep{2013A&A...553L...7I,2016ApJ...826...71I,2017A&A...599A..71D}. 

More work is needed in this direction. On one hand, high-redshift SN Ia host galaxy spectroscopy in the near-infrared is needed to extend the Hubble diagram above 0.5. We have shown that higher-redshift galaxies are metal-poorer than nearby galaxies, so including those objects in an analysis similar to what we presented in this work would expand the dynamic range of abundances and therefore allow for steeper relations with Hubble residuals. In the third part of our project (see section \ref{Section1}) we will perform a similar analysis to that presented here but extending the redshift range beyond z$>$0.5, where differences among different cosmological models start to become evident, and also where the effect of the chemical enrichment evolution of the Universe may have a strong role.

On the other hand, performing local instead of global measurements of the oxygen abundance would be more reliable given that small- and large-scale galactic processes modify the abundance at different locations of galaxies. This effect would be more significant for those SNe Ia that explode further from the galaxy core. Fortunately, recent and future works would be able to provide such measurements through Integral Field Spectroscopy of galaxies at different redshifts \citep{2016MNRAS.455.4087G}.

\section*{Acknowledgements}
This work has been supported by DGICYT/MINECO grants AYA2010-21887-C04-02 and AYA2013-47742-C4-4-P. M.E. M-R acknowledges financial support from the Ministerio de Econom\'ia y Competitividad (MINECO) through two grants for research visits to the Australian Astronomical Observatory (AAO) and to the Department of Astronomy of Universidad de Chile where this work was partially done. L.G. was supported in part by the US National Science Foundation under Grant AST-1311862. This work use Sloan Digital Sky Survey (SDSS) data. Funding for the SDSS and SDSS-II was provided by the Alfred P. Sloan Foundation, the Participating Institutions, the National Science Foundation, the U.S. Department of Energy, the National Aeronautics and Space Administration, the Japanese Monbukagakusho, the Max Planck Society, and the Higher Education Funding Council for England. The SDSS was managed by the Astrophysical Research Consortium for the Participating Institutions. IRAF is distributed by NOAO which is operated by AURA Inc., under cooperative agreement with NSF.

\bibliographystyle{mn2e}
\bibliography{SDSS_OH}

\begin{thebibliography}{102}
\expandafter\ifx\csname natexlab\endcsname\relax\def\natexlab#1{#1}\fi

\bibitem[{{Alam} {et~al}\mbox{.}(2015){Alam}, {Albareti}, {Allende Prieto},
  {Anders}, {Anderson}, {Anderton}, {Andrews}, {Armengaud}, {Aubourg},
  {Bailey}, \& et~al.}]{2015ApJS..219...12A}
{Alam} S. {et~al.}, 2015, \apjs, 219, 12

\bibitem[{{Asari} {et~al}\mbox{.}(2007){Asari}, {Cid Fernandes},
  {Stasi{\'n}ska}, {Torres-Papaqui}, {Mateus}, {Sodr{\'e}}, {Schoenell}, \&
  {Gomes}}]{2007MNRAS.381..263A}
{Asari} N.~V., {Cid Fernandes} R., {Stasi{\'n}ska} G., {Torres-Papaqui} J.~P.,
  {Mateus} A., {Sodr{\'e}} L., {Schoenell} W., {Gomes} J.~M., 2007, \mnras,
  381, 263

\bibitem[{{Asplund} {et~al}\mbox{.}(2009){Asplund}, {Grevesse}, {Sauval}, \&
  {Scott}}]{2009ARA&A..47..481A}
{Asplund} M., {Grevesse} N., {Sauval} A.~J., {Scott} P., 2009, \araa, 47, 481

\bibitem[{{Baldwin}, {Phillips} \& {Terlevich}(1981){Baldwin}, {Phillips}, \&
  {Terlevich}}]{1981PASP...93....5B}
{Baldwin} J.~A., {Phillips} M.~M., {Terlevich} R., 1981, \pasp, 93, 5

\bibitem[{{Barris} \& {Tonry}(2004)}]{2004ApJ...613L..21B}
{Barris} B.~J., {Tonry} J.~L., 2004, \apjl, 613, L21

\bibitem[{{Betoule} {et~al}\mbox{.}(2014){Betoule}, {Kessler}, {Guy}, {Mosher},
  {Hardin}, {Biswas}, {Astier}, {El-Hage}, {Konig}, {Kuhlmann}, {Marriner},
  {Pain}, {Regnault}, {Balland}, {Bassett}, {Brown}, {Campbell}, {Carlberg},
  {Cellier-Holzem}, {Cinabro}, {Conley}, {D'Andrea}, {DePoy}, {Doi}, {Ellis},
  {Fabbro}, {Filippenko}, {Foley}, {Frieman}, {Fouchez}, {Galbany}, {Goobar},
  {Gupta}, {Hill}, {Hlozek}, {Hogan}, {Hook}, {Howell}, {Jha}, {Le Guillou},
  {Leloudas}, {Lidman}, {Marshall}, {M{\"o}ller}, {Mour{\~a}o}, {Neveu},
  {Nichol}, {Olmstead}, {Palanque-Delabrouille}, {Perlmutter}, {Prieto},
  {Pritchet}, {Richmond}, {Riess}, {Ruhlmann-Kleider}, {Sako}, {Schahmaneche},
  {Schneider}, {Smith}, {Sollerman}, {Sullivan}, {Walton}, \&
  {Wheeler}}]{2014A&A...568A..22B}
{Betoule} M. {et~al.}, 2014, \aap, 568, A22

\bibitem[{{Bravo} {et~al}\mbox{.}(2010){Bravo}, {Dom{\'{\i}}nguez}, {Badenes},
  {Piersanti}, \& {Straniero}}]{2010ApJ...711L..66B}
{Bravo} E., {Dom{\'{\i}}nguez} I., {Badenes} C., {Piersanti} L., {Straniero}
  O., 2010, \apjl, 711, L66

\bibitem[{{Bruzual} \& {Charlot}(2003)}]{2003MNRAS.344.1000B}
{Bruzual} G., {Charlot} S., 2003, \mnras, 344, 1000

\bibitem[{{Burns} {et~al}\mbox{.}(2011){Burns}, {Stritzinger}, {Phillips},
  {Kattner}, {Persson}, {Madore}, {Freedman}, {Boldt}, {Campillay},
  {Contreras}, {Folatelli}, {Gonzalez}, {Krzeminski}, {Morrell}, {Salgado}, \&
  {Suntzeff}}]{2011AJ....141...19B}
{Burns} C.~R. {et~al.}, 2011, \aj, 141, 19

\bibitem[{{Campbell}, {Fraser} \& {Gilmore}(2016){Campbell}, {Fraser}, \&
  {Gilmore}}]{2016MNRAS.457.3470C}
{Campbell} H., {Fraser} M., {Gilmore} G., 2016, \mnras, 457, 3470

\bibitem[{{Cardelli}, {Clayton} \& {Mathis}(1989){Cardelli}, {Clayton}, \&
  {Mathis}}]{1989ApJ...345..245C}
{Cardelli} J.~A., {Clayton} G.~C., {Mathis} J.~S., 1989, \apj, 345, 245

\bibitem[{{Chabrier}(2003)}]{2003PASP..115..763C}
{Chabrier} G., 2003, \pasp, 115, 763

\bibitem[{{Childress} {et~al}\mbox{.}(2013){Childress}, {Aldering},
  {Antilogus}, {Aragon}, {Bailey}, {Baltay}, {Bongard}, {Buton}, {Canto},
  {Cellier-Holzem}, {Chotard}, {Copin}, {Fakhouri}, {Gangler}, {Guy}, {Hsiao},
  {Kerschhaggl}, {Kim}, {Kowalski}, {Loken}, {Nugent}, {Paech}, {Pain},
  {Pecontal}, {Pereira}, {Perlmutter}, {Rabinowitz}, {Rigault}, {Runge},
  {Scalzo}, {Smadja}, {Tao}, {Thomas}, {Weaver}, \& {Wu}}]{2013ApJ...770..108C}
{Childress} M. {et~al.}, 2013, \apj, 770, 108

\bibitem[{{Cid Fernandes} {et~al}\mbox{.}(2005){Cid Fernandes}, {Mateus},
  {Sodr{\'e}}, {Stasi{\'n}ska}, \& {Gomes}}]{2005MNRAS.358..363C}
{Cid Fernandes} R., {Mateus} A., {Sodr{\'e}} L., {Stasi{\'n}ska} G., {Gomes}
  J.~M., 2005, \mnras, 358, 363

\bibitem[{{Cid Fernandes} {et~al}\mbox{.}(2009){Cid Fernandes}, {Schoenell},
  {Gomes}, {Asari}, {Schlickmann}, {Mateus}, {Stasinska}, {Sodr{\'e}},
  {Torres-Papaqui}, \& {Seagal Collaboration}}]{2009RMxAC..35..127C}
{Cid Fernandes} R. {et~al.}, 2009, in Revista Mexicana de Astronomia y
  Astrofisica, vol.~27, Vol.~35, Revista Mexicana de Astronomia y Astrofisica
  Conference Series, pp. 127--132

\bibitem[{{Conley} {et~al}\mbox{.}(2011){Conley}, {Guy}, {Sullivan},
  {Regnault}, {Astier}, {Balland}, {Basa}, {Carlberg}, {Fouchez}, {Hardin},
  {Hook}, {Howell}, {Pain}, {Palanque-Delabrouille}, {Perrett}, {Pritchet},
  {Rich}, {Ruhlmann-Kleider}, {Balam}, {Baumont}, {Ellis}, {Fabbro},
  {Fakhouri}, {Fourmanoit}, {Gonz{\'a}lez-Gait{\'a}n}, {Graham}, {Hudson},
  {Hsiao}, {Kronborg}, {Lidman}, {Mourao}, {Neill}, {Perlmutter}, {Ripoche},
  {Suzuki}, \& {Walker}}]{2011ApJS..192....1C}
{Conley} A. {et~al.}, 2011, \apjs, 192, 1

\bibitem[{{Conley} {et~al}\mbox{.}(2008){Conley}, {Sullivan}, {Hsiao}, {Guy},
  {Astier}, {Balam}, {Balland}, {Basa}, {Carlberg}, {Fouchez}, {Hardin},
  {Howell}, {Hook}, {Pain}, {Perrett}, {Pritchet}, \&
  {Regnault}}]{2008ApJ...681..482C}
{Conley} A. {et~al.}, 2008, \apj, 681, 482

\bibitem[{{D'Andrea} {et~al}\mbox{.}(2011){D'Andrea}, {Gupta}, {Sako},
  {Morris}, {Nichol}, {Brown}, {Campbell}, {Olmstead}, {Frieman}, {Garnavich},
  {Jha}, {Kessler}, {Lampeitl}, {Marriner}, {Schneider}, \&
  {Smith}}]{2011ApJ...743..172D}
{D'Andrea} C.~B. {et~al.}, 2011, \apj, 743, 172

\bibitem[{{Dawson} {et~al}\mbox{.}(2013){Dawson}, {Schlegel}, {Ahn},
  {Anderson}, {Aubourg}, {Bailey}, {Barkhouser}, {Bautista}, {Beifiori},
  {Berlind}, {Bhardwaj}, {Bizyaev}, {Blake}, {Blanton}, {Blomqvist}, {Bolton},
  {Borde}, {Bovy}, {Brandt}, {Brewington}, {Brinkmann}, {Brown}, {Brownstein},
  {Bundy}, {Busca}, {Carithers}, {Carnero}, {Carr}, {Chen}, {Comparat},
  {Connolly}, {Cope}, {Croft}, {Cuesta}, {da Costa}, {Davenport}, {Delubac},
  {de Putter}, {Dhital}, {Ealet}, {Ebelke}, {Eisenstein}, {Escoffier}, {Fan},
  {Filiz Ak}, {Finley}, {Font-Ribera}, {G{\'e}nova-Santos}, {Gunn}, {Guo},
  {Haggard}, {Hall}, {Hamilton}, {Harris}, {Harris}, {Ho}, {Hogg}, {Holder},
  {Honscheid}, {Huehnerhoff}, {Jordan}, {Jordan}, {Kauffmann}, {Kazin},
  {Kirkby}, {Klaene}, {Kneib}, {Le Goff}, {Lee}, {Long}, {Loomis}, {Lundgren},
  {Lupton}, {Maia}, {Makler}, {Malanushenko}, {Malanushenko}, {Mandelbaum},
  {Manera}, {Maraston}, {Margala}, {Masters}, {McBride}, {McDonald}, {McGreer},
  {McMahon}, {Mena}, {Miralda-Escud{\'e}}, {Montero-Dorta}, {Montesano},
  {Muna}, {Myers}, {Naugle}, {Nichol}, {Noterdaeme}, {Nuza}, {Olmstead},
  {Oravetz}, {Oravetz}, {Owen}, {Padmanabhan}, {Palanque-Delabrouille}, {Pan},
  {Parejko}, {P{\^a}ris}, {Percival}, {P{\'e}rez-Fournon},
  {P{\'e}rez-R{\`a}fols}, {Petitjean}, {Pfaffenberger}, {Pforr}, {Pieri},
  {Prada}, {Price-Whelan}, {Raddick}, {Rebolo}, {Rich}, {Richards}, {Rockosi},
  {Roe}, {Ross}, {Ross}, {Rossi}, {Rubi{\~n}o-Martin}, {Samushia},
  {S{\'a}nchez}, {Sayres}, {Schmidt}, {Schneider}, {Sc{\'o}ccola}, {Seo},
  {Shelden}, {Sheldon}, {Shen}, {Shu}, {Slosar}, {Smee}, {Snedden}, {Stauffer},
  {Steele}, {Strauss}, {Streblyanska}, {Suzuki}, {Swanson}, {Tal}, {Tanaka},
  {Thomas}, {Tinker}, {Tojeiro}, {Tremonti}, {Vargas Maga{\~n}a}, {Verde},
  {Viel}, {Wake}, {Watson}, {Weaver}, {Weinberg}, {Weiner}, {West}, {White},
  {Wood-Vasey}, {Yeche}, {Zehavi}, {Zhao}, \& {Zheng}}]{2013AJ....145...10D}
{Dawson} K.~S. {et~al.}, 2013, \aj, 145, 10

\bibitem[{{de Jaeger} {et~al}\mbox{.}(2017){de Jaeger}, {Galbany},
  {Filippenko}, {Gonz{\'a}lez-Gait{\'a}n}, {Yasuda}, {Maeda}, {Tanaka},
  {Morokuma}, {Moriya}, {Tominaga}, {Nomoto}, {Komiyama}, {Anderson}, {Brink},
  {Carlberg}, {Folatelli}, {Hamuy}, {Pignata}, \&
  {Zheng}}]{2017MNRAS.472.4233D}
{de Jaeger} T. {et~al.}, 2017, \mnras, 472, 4233

\bibitem[{{Duarte Puertas} {et~al}\mbox{.}(2017){Duarte Puertas}, {Vilchez},
  {Iglesias-P{\'a}ramo}, {Kehrig}, {P{\'e}rez-Montero}, \&
  {Rosales-Ortega}}]{2017A&A...599A..71D}
{Duarte Puertas} S., {Vilchez} J.~M., {Iglesias-P{\'a}ramo} J., {Kehrig} C.,
  {P{\'e}rez-Montero} E., {Rosales-Ortega} F.~F., 2017, \aap, 599, A71

\bibitem[{{Foreman-Mackey} {et~al}\mbox{.}(2013){Foreman-Mackey}, {Hogg},
  {Lang}, \& {Goodman}}]{2013PASP..125..306F}
{Foreman-Mackey} D., {Hogg} D.~W., {Lang} D., {Goodman} J., 2013, \pasp, 125,
  306

\bibitem[{{Frieman} {et~al}\mbox{.}(2008){Frieman}, {Bassett}, {Becker},
  {Choi}, {Cinabro}, {DeJongh}, {Depoy}, {Dilday}, {Doi}, {Garnavich}, {Hogan},
  {Holtzman}, {Im}, {Jha}, {Kessler}, {Konishi}, {Lampeitl}, {Marriner},
  {Marshall}, {McGinnis}, {Miknaitis}, {Nichol}, {Prieto}, {Riess}, {Richmond},
  {Romani}, {Sako}, {Schneider}, {Smith}, {Takanashi}, {Tokita}, {van der
  Heyden}, {Yasuda}, {Zheng}, {Adelman-McCarthy}, {Annis}, {Assef},
  {Barentine}, {Bender}, {Blandford}, {Boroski}, {Bremer}, {Brewington},
  {Collins}, {Crotts}, {Dembicky}, {Eastman}, {Edge}, {Edmondson}, {Elson},
  {Eyler}, {Filippenko}, {Foley}, {Frank}, {Goobar}, {Gueth}, {Gunn},
  {Harvanek}, {Hopp}, {Ihara}, {Ivezi{\'c}}, {Kahn}, {Kaplan}, {Kent},
  {Ketzeback}, {Kleinman}, {Kollatschny}, {Kron}, {Krzesi{\'n}ski}, {Lamenti},
  {Leloudas}, {Lin}, {Long}, {Lucey}, {Lupton}, {Malanushenko}, {Malanushenko},
  {McMillan}, {Mendez}, {Morgan}, {Morokuma}, {Nitta}, {Ostman}, {Pan},
  {Rockosi}, {Romer}, {Ruiz-Lapuente}, {Saurage}, {Schlesinger}, {Snedden},
  {Sollerman}, {Stoughton}, {Stritzinger}, {Subba Rao}, {Tucker}, {Vaisanen},
  {Watson}, {Watters}, {Wheeler}, {Yanny}, \& {York}}]{2008AJ....135..338F}
{Frieman} J.~A. {et~al.}, 2008, \aj, 135, 338

\bibitem[{{Galbany} {et~al}\mbox{.}(2016{\natexlab{a}}){Galbany}, {Anderson},
  {Rosales-Ortega}, {Kuncarayakti}, {Kr{\"u}hler}, {S{\'a}nchez},
  {Falc{\'o}n-Barroso}, {P{\'e}rez}, {Maureira}, {Hamuy},
  {Gonz{\'a}lez-Gait{\'a}n}, {F{\"o}rster}, \& {Moral}}]{2016MNRAS.455.4087G}
{Galbany} L. {et~al.}, 2016{\natexlab{a}}, \mnras, 455, 4087

\bibitem[{{Galbany} {et~al}\mbox{.}(2012){Galbany}, {Miquel}, {{\"O}stman},
  {Brown}, {Cinabro}, {D'Andrea}, {Frieman}, {Jha}, {Marriner}, {Nichol},
  {Nordin}, {Olmstead}, {Sako}, {Schneider}, {Smith}, {Sollerman}, {Pan},
  {Snedden}, {Bizyaev}, {Brewington}, {Malanushenko}, {Malanushenko},
  {Oravetz}, {Simmons}, \& {Shelden}}]{2012ApJ...755..125G}
{Galbany} L. {et~al.}, 2012, \apj, 755, 125

\bibitem[{{Galbany} {et~al}\mbox{.}(2014){Galbany}, {Stanishev}, {Mour{\~a}o},
  {Rodrigues}, {Flores}, {Garc{\'{\i}}a-Benito}, {Mast}, {Mendoza},
  {S{\'a}nchez}, {Badenes}, {Barrera-Ballesteros}, {Bland-Hawthorn},
  {Falc{\'o}n-Barroso}, {Garc{\'{\i}}a-Lorenzo}, {Gomes}, {Gonz{\'a}lez
  Delgado}, {Kehrig}, {Lyubenova}, {L{\'o}pez-S{\'a}nchez}, {de
  Lorenzo-C{\'a}ceres}, {Marino}, {Meidt}, {Moll{\'a}}, {Papaderos},
  {P{\'e}rez-Torres}, {Rosales-Ortega}, \& {van de Ven}}]{2014A&A...572A..38G}
{Galbany} L. {et~al.}, 2014, \aap, 572, A38

\bibitem[{{Galbany} {et~al}\mbox{.}(2016{\natexlab{b}}){Galbany}, {Stanishev},
  {Mour{\~a}o}, {Rodrigues}, {Flores}, {Walcher}, {S{\'a}nchez},
  {Garc{\'{\i}}a-Benito}, {Mast}, {Badenes}, {Gonz{\'a}lez Delgado}, {Kehrig},
  {Lyubenova}, {Marino}, {Moll{\'a}}, {Meidt}, {P{\'e}rez}, {van de Ven}, \&
  {V{\'{\i}}lchez}}]{2016A&A...591A..48G}
{Galbany} L. {et~al.}, 2016{\natexlab{b}}, \aap, 591, A48

\bibitem[{{Gallagher} {et~al}\mbox{.}(2008){Gallagher}, {Garnavich},
  {Caldwell}, {Kirshner}, {Jha}, {Li}, {Ganeshalingam}, \&
  {Filippenko}}]{2008ApJ...685..752G}
{Gallagher} J.~S., {Garnavich} P.~M., {Caldwell} N., {Kirshner} R.~P., {Jha}
  S.~W., {Li} W., {Ganeshalingam} M., {Filippenko} A.~V., 2008, \apj, 685, 752

\bibitem[{{Garc{\'{\i}}a-Rojas} {et~al}\mbox{.}(2004){Garc{\'{\i}}a-Rojas},
  {Esteban}, {Peimbert}, {Rodr{\'{\i}}guez}, {Ruiz}, \&
  {Peimbert}}]{2004ApJS..153..501G}
{Garc{\'{\i}}a-Rojas} J., {Esteban} C., {Peimbert} M., {Rodr{\'{\i}}guez} M.,
  {Ruiz} M.~T., {Peimbert} A., 2004, \apjs, 153, 501

\bibitem[{{Garnett}(1992)}]{1992AJ....103.1330G}
{Garnett} D.~R., 1992, \aj, 103, 1330

\bibitem[{{Gunn} {et~al}\mbox{.}(1998){Gunn}, {Carr}, {Rockosi}, {Sekiguchi},
  {Berry}, {Elms}, {de Haas}, {Ivezi{\'c}}, {Knapp}, {Lupton}, {Pauls},
  {Simcoe}, {Hirsch}, {Sanford}, {Wang}, {York}, {Harris}, {Annis}, {Bartozek},
  {Boroski}, {Bakken}, {Haldeman}, {Kent}, {Holm}, {Holmgren}, {Petravick},
  {Prosapio}, {Rechenmacher}, {Doi}, {Fukugita}, {Shimasaku}, {Okada}, {Hull},
  {Siegmund}, {Mannery}, {Blouke}, {Heidtman}, {Schneider}, {Lucinio}, \&
  {Brinkman}}]{1998AJ....116.3040G}
{Gunn} J.~E. {et~al.}, 1998, \aj, 116, 3040

\bibitem[{{Gunn} {et~al}\mbox{.}(2006){Gunn}, {Siegmund}, {Mannery}, {Owen},
  {Hull}, {Leger}, {Carey}, {Knapp}, {York}, {Boroski}, {Kent}, {Lupton},
  {Rockosi}, {Evans}, {Waddell}, {Anderson}, {Annis}, {Barentine}, {Bartoszek},
  {Bastian}, {Bracker}, {Brewington}, {Briegel}, {Brinkmann}, {Brown}, {Carr},
  {Czarapata}, {Drennan}, {Dombeck}, {Federwitz}, {Gillespie}, {Gonzales},
  {Hansen}, {Harvanek}, {Hayes}, {Jordan}, {Kinney}, {Klaene}, {Kleinman},
  {Kron}, {Kresinski}, {Lee}, {Limmongkol}, {Lindenmeyer}, {Long}, {Loomis},
  {McGehee}, {Mantsch}, {Neilsen}, {Neswold}, {Newman}, {Nitta}, {Peoples},
  {Pier}, {Prieto}, {Prosapio}, {Rivetta}, {Schneider}, {Snedden}, \&
  {Wang}}]{2006AJ....131.2332G}
{Gunn} J.~E. {et~al.}, 2006, \aj, 131, 2332

\bibitem[{{Gupta} {et~al}\mbox{.}(2011){Gupta}, {D'Andrea}, {Sako}, {Conroy},
  {Smith}, {Bassett}, {Frieman}, {Garnavich}, {Jha}, {Kessler}, {Lampeitl},
  {Marriner}, {Nichol}, \& {Schneider}}]{2011ApJ...740...92G}
{Gupta} R.~R. {et~al.}, 2011, \apj, 740, 92

\bibitem[{{Guy} {et~al}\mbox{.}(2007){Guy}, {Astier}, {Baumont}, {Hardin},
  {Pain}, {Regnault}, {Basa}, {Carlberg}, {Conley}, {Fabbro}, {Fouchez},
  {Hook}, {Howell}, {Perrett}, {Pritchet}, {Rich}, {Sullivan}, {Antilogus},
  {Aubourg}, {Bazin}, {Bronder}, {Filiol}, {Palanque-Delabrouille}, {Ripoche},
  \& {Ruhlmann-Kleider}}]{2007A&A...466...11G}
{Guy} J. {et~al.}, 2007, \aap, 466, 11

\bibitem[{{Guy} {et~al}\mbox{.}(2005){Guy}, {Astier}, {Nobili}, {Regnault}, \&
  {Pain}}]{2005A&A...443..781G}
{Guy} J., {Astier} P., {Nobili} S., {Regnault} N., {Pain} R., 2005, \aap, 443,
  781

\bibitem[{{Henry} \& {Worthey}(1999)}]{1999PASP..111..919H}
{Henry} R.~B.~C., {Worthey} G., 1999, \pasp, 111, 919

\bibitem[{{Hicken} {et~al}\mbox{.}(2009){Hicken}, {Challis}, {Jha}, {Kirshner},
  {Matheson}, {Modjaz}, {Rest}, {Wood-Vasey}, {Bakos}, {Barton}, {Berlind},
  {Bragg}, {Brice{\~n}o}, {Brown}, {Caldwell}, {Calkins}, {Cho}, {Ciupik},
  {Contreras}, {Dendy}, {Dosaj}, {Durham}, {Eriksen}, {Esquerdo}, {Everett},
  {Falco}, {Fernandez}, {Gaba}, {Garnavich}, {Graves}, {Green}, {Groner},
  {Hergenrother}, {Holman}, {Hradecky}, {Huchra}, {Hutchison}, {Jerius},
  {Jordan}, {Kilgard}, {Krauss}, {Luhman}, {Macri}, {Marrone}, {McDowell},
  {McIntosh}, {McNamara}, {Megeath}, {Mochejska}, {Munoz}, {Muzerolle},
  {Naranjo}, {Narayan}, {Pahre}, {Peters}, {Peterson}, {Rines}, {Ripman},
  {Roussanova}, {Schild}, {Sicilia-Aguilar}, {Sokoloski}, {Smalley}, {Smith},
  {Spahr}, {Stanek}, {Barmby}, {Blondin}, {Stubbs}, {Szentgyorgyi}, {Torres},
  {Vaz}, {Vikhlinin}, {Wang}, {Westover}, {Woods}, \&
  {Zhao}}]{2009ApJ...700..331H}
{Hicken} M. {et~al.}, 2009, \apj, 700, 331

\bibitem[{{Howell} {et~al}\mbox{.}(2009){Howell}, {Sullivan}, {Brown},
  {Conley}, {Le Borgne}, {Hsiao}, {Astier}, {Balam}, {Balland}, {Basa},
  {Carlberg}, {Fouchez}, {Guy}, {Hardin}, {Hook}, {Pain}, {Perrett},
  {Pritchet}, {Regnault}, {Baumont}, {LeDu}, {Lidman}, {Perlmutter}, {Suzuki},
  {Walker}, \& {Wheeler}}]{2009ApJ...691..661H}
{Howell} D.~A. {et~al.}, 2009, \apj, 691, 661

\bibitem[{{Iglesias-P{\'a}ramo} {et~al}\mbox{.}(2013){Iglesias-P{\'a}ramo},
  {V{\'{\i}}lchez}, {Galbany}, {S{\'a}nchez}, {Rosales-Ortega}, {Mast},
  {Garc{\'{\i}}a-Benito}, {Husemann}, {Aguerri}, {Alves}, {Bekerait{\'e}},
  {Bland-Hawthorn}, {Catal{\'a}n-Torrecilla}, {de Amorim}, {de
  Lorenzo-C{\'a}ceres}, {Ellis}, {Falc{\'o}n-Barroso}, {Flores}, {Florido},
  {Gallazzi}, {Gomes}, {Gonz{\'a}lez Delgado}, {Haines},
  {Hern{\'a}ndez-Fern{\'a}ndez}, {Kehrig}, {L{\'o}pez-S{\'a}nchez},
  {Lyubenova}, {Marino}, {Moll{\'a}}, {Monreal-Ibero}, {Mour{\~a}o},
  {Papaderos}, {Rodrigues}, {S{\'a}nchez-Bl{\'a}zquez}, {Spekkens},
  {Stanishev}, {van de Ven}, {Walcher}, {Wisotzki}, {Zibetti}, \&
  {Ziegler}}]{2013A&A...553L...7I}
{Iglesias-P{\'a}ramo} J. {et~al.}, 2013, \aap, 553, L7

\bibitem[{{Iglesias-P{\'a}ramo} {et~al}\mbox{.}(2016){Iglesias-P{\'a}ramo},
  {V{\'{\i}}lchez}, {Rosales-Ortega}, {S{\'a}nchez}, {Duarte Puertas},
  {Petropoulou}, {Gil de Paz}, {Galbany}, {Moll{\'a}},
  {Catal{\'a}n-Torrecilla}, {Castillo Morales}, {Mast}, {Husemann},
  {Garc{\'{\i}}a-Benito}, {Mendoza}, {Kehrig}, {P{\'e}rez-Montero},
  {Papaderos}, {Gomes}, {Walcher}, {Gonz{\'a}lez Delgado}, {Marino},
  {L{\'o}pez-S{\'a}nchez}, {Ziegler}, {Flores}, \&
  {Alves}}]{2016ApJ...826...71I}
{Iglesias-P{\'a}ramo} J. {et~al.}, 2016, \apj, 826, 71

\bibitem[{{Jha}, {Riess} \& {Kirshner}(2007){Jha}, {Riess}, \&
  {Kirshner}}]{2007ApJ...659..122J}
{Jha} S., {Riess} A.~G., {Kirshner} R.~P., 2007, \apj, 659, 122

\bibitem[{{Johansson} {et~al}\mbox{.}(2013){Johansson}, {Thomas}, {Pforr},
  {Maraston}, {Nichol}, {Smith}, {Lampeitl}, {Beifiori}, {Gupta}, \&
  {Schneider}}]{2013MNRAS.435.1680J}
{Johansson} J. {et~al.}, 2013, \mnras, 435, 1680

\bibitem[{{Kasen}, {R{\"o}pke} \& {Woosley}(2009){Kasen}, {R{\"o}pke}, \&
  {Woosley}}]{kasen09}
{Kasen} D., {R{\"o}pke} F.~K., {Woosley} S.~E., 2009, Nature, 460, 869

\bibitem[{{Kauffmann} {et~al}\mbox{.}(2003){Kauffmann}, {Heckman}, {Tremonti},
  {Brinchmann}, {Charlot}, {White}, {Ridgway}, {Brinkmann}, {Fukugita}, {Hall},
  {Ivezi{\'c}}, {Richards}, \& {Schneider}}]{2003MNRAS.346.1055K}
{Kauffmann} G. {et~al.}, 2003, \mnras, 346, 1055

\bibitem[{{Kelly} {et~al}\mbox{.}(2010){Kelly}, {Hicken}, {Burke}, {Mandel}, \&
  {Kirshner}}]{2010ApJ...715..743K}
{Kelly} P.~L., {Hicken} M., {Burke} D.~L., {Mandel} K.~S., {Kirshner} R.~P.,
  2010, \apj, 715, 743

\bibitem[{{Kewley} \& {Dopita}(2002)}]{2002ApJS..142...35K}
{Kewley} L.~J., {Dopita} M.~A., 2002, \apjs, 142, 35

\bibitem[{{Kewley} {et~al}\mbox{.}(2001){Kewley}, {Dopita}, {Sutherland},
  {Heisler}, \& {Trevena}}]{2001ApJ...556..121K}
{Kewley} L.~J., {Dopita} M.~A., {Sutherland} R.~S., {Heisler} C.~A., {Trevena}
  J., 2001, \apj, 556, 121

\bibitem[{{Kewley} \& {Ellison}(2008)}]{2008ApJ...681.1183K}
{Kewley} L.~J., {Ellison} S.~L., 2008, \apj, 681, 1183

\bibitem[{{Kewley} {et~al}\mbox{.}(2006){Kewley}, {Groves}, {Kauffmann}, \&
  {Heckman}}]{2006MNRAS.372..961K}
{Kewley} L.~J., {Groves} B., {Kauffmann} G., {Heckman} T., 2006, \mnras, 372,
  961

\bibitem[{{Kobulnicky} \& {Kewley}(2004)}]{2004ApJ...617..240K}
{Kobulnicky} H.~A., {Kewley} L.~J., 2004, \apj, 617, 240

\bibitem[{{Lampeitl} {et~al}\mbox{.}(2010){Lampeitl}, {Smith}, {Nichol},
  {Bassett}, {Cinabro}, {Dilday}, {Foley}, {Frieman}, {Garnavich}, {Goobar},
  {Im}, {Jha}, {Marriner}, {Miquel}, {Nordin}, {{\"O}stman}, {Riess}, {Sako},
  {Schneider}, {Sollerman}, \& {Stritzinger}}]{2010ApJ...722..566L}
{Lampeitl} H. {et~al.}, 2010, \apj, 722, 566

\bibitem[{{Lara-L{\'o}pez}, {L{\'o}pez-S{\'a}nchez} \&
  {Hopkins}(2013){Lara-L{\'o}pez}, {L{\'o}pez-S{\'a}nchez}, \&
  {Hopkins}}]{2013ApJ...764..178L}
{Lara-L{\'o}pez} M.~A., {L{\'o}pez-S{\'a}nchez} {\'A}.~R., {Hopkins} A.~M.,
  2013, \apj, 764, 178

\bibitem[{{L{\'o}pez-S{\'a}nchez} {et~al}\mbox{.}(2012){L{\'o}pez-S{\'a}nchez},
  {Dopita}, {Kewley}, {Zahid}, {Nicholls}, \&
  {Scharw{\"a}chter}}]{2012MNRAS.426.2630L}
{L{\'o}pez-S{\'a}nchez} {\'A}.~R., {Dopita} M.~A., {Kewley} L.~J., {Zahid}
  H.~J., {Nicholls} D.~C., {Scharw{\"a}chter} J., 2012, \mnras, 426, 2630

\bibitem[{{L{\'o}pez-S{\'a}nchez} \& {Esteban}(2009)}]{2009A&A...508..615L}
{L{\'o}pez-S{\'a}nchez} A.~R., {Esteban} C., 2009, \aap, 508, 615

\bibitem[{{L{\'o}pez-S{\'a}nchez} \& {Esteban}(2010)}]{2010A&A...517A..85L}
{L{\'o}pez-S{\'a}nchez} {\'A}.~R., {Esteban} C., 2010, \aap, 517, A85

\bibitem[{{L{\'o}pez-S{\'a}nchez} {et~al}\mbox{.}(2015){L{\'o}pez-S{\'a}nchez},
  {Westmeier}, {Esteban}, \& {Koribalski}}]{2015MNRAS.450.3381L}
{L{\'o}pez-S{\'a}nchez} {\'A}.~R., {Westmeier} T., {Esteban} C., {Koribalski}
  B.~S., 2015, \mnras, 450, 3381

\bibitem[{{Maoz}, {Mannucci} \& {Nelemans}(2014){Maoz}, {Mannucci}, \&
  {Nelemans}}]{2014ARA&A..52..107M}
{Maoz} D., {Mannucci} F., {Nelemans} G., 2014, \araa, 52, 107

\bibitem[{{Marigo} \& {Girardi}(2007)}]{2007A&A...469..239M}
{Marigo} P., {Girardi} L., 2007, \aap, 469, 239

\bibitem[{{Marigo} {et~al}\mbox{.}(2008){Marigo}, {Girardi}, {Bressan},
  {Groenewegen}, {Silva}, \& {Granato}}]{2008A&A...482..883M}
{Marigo} P., {Girardi} L., {Bressan} A., {Groenewegen} M.~A.~T., {Silva} L.,
  {Granato} G.~L., 2008, \aap, 482, 883

\bibitem[{{Marino} {et~al}\mbox{.}(2013){Marino}, {Rosales-Ortega},
  {S{\'a}nchez}, {Gil de Paz}, {V{\'{\i}}lchez}, {Miralles-Caballero},
  {Kehrig}, {P{\'e}rez-Montero}, {Stanishev}, {Iglesias-P{\'a}ramo},
  {D{\'{\i}}az}, {Castillo-Morales}, {Kennicutt}, {L{\'o}pez-S{\'a}nchez},
  {Galbany}, {Garc{\'{\i}}a-Benito}, {Mast}, {Mendez-Abreu}, {Monreal-Ibero},
  {Husemann}, {Walcher}, {Garc{\'{\i}}a-Lorenzo}, {Masegosa}, {Del Olmo
  Orozco}, {Mour{\~a}o}, {Ziegler}, {Moll{\'a}}, {Papaderos},
  {S{\'a}nchez-Bl{\'a}zquez}, {Gonz{\'a}lez Delgado}, {Falc{\'o}n-Barroso},
  {Roth}, {van de Ven}, \& {Califa Team}}]{2013A&A...559A.114M}
{Marino} R.~A. {et~al.}, 2013, \aap, 559, A114

\bibitem[{{Mateus} {et~al}\mbox{.}(2006){Mateus}, {Sodr{\'e}}, {Cid Fernandes},
  {Stasi{\'n}ska}, {Schoenell}, \& {Gomes}}]{2006MNRAS.370..721M}
{Mateus} A., {Sodr{\'e}} L., {Cid Fernandes} R., {Stasi{\'n}ska} G.,
  {Schoenell} W., {Gomes} J.~M., 2006, \mnras, 370, 721

\bibitem[{{Mazzarella} \& {Boroson}(1993)}]{1993ApJS...85...27M}
{Mazzarella} J.~M., {Boroson} T.~A., 1993, \apjs, 85, 27

\bibitem[{{Moll{\'a}} \& {D{\'{\i}}az}(2005)}]{2005MNRAS.358..521M}
{Moll{\'a}} M., {D{\'{\i}}az} A.~I., 2005, \mnras, 358, 521

\bibitem[{{Moreno-Raya} {et~al}\mbox{.}(2016{\natexlab{a}}){Moreno-Raya},
  {L{\'o}pez-S{\'a}nchez}, {Moll{\'a}}, {Galbany}, {V{\'{\i}}lchez}, \&
  {Carnero}}]{2016MNRAS.462.1281M}
{Moreno-Raya} M.~E., {L{\'o}pez-S{\'a}nchez} {\'A}.~R., {Moll{\'a}} M.,
  {Galbany} L., {V{\'{\i}}lchez} J.~M., {Carnero} A., 2016{\natexlab{a}},
  \mnras, 462, 1281

\bibitem[{{Moreno-Raya} {et~al}\mbox{.}(2016{\natexlab{b}}){Moreno-Raya},
  {Moll{\'a}}, {L{\'o}pez-S{\'a}nchez}, {Galbany}, {V{\'{\i}}lchez}, {Carnero
  Rosell}, \& {Dom{\'{\i}}nguez}}]{2016ApJ...818L..19M}
{Moreno-Raya} M.~E., {Moll{\'a}} M., {L{\'o}pez-S{\'a}nchez} {\'A}.~R.,
  {Galbany} L., {V{\'{\i}}lchez} J.~M., {Carnero Rosell} A., {Dom{\'{\i}}nguez}
  I., 2016{\natexlab{b}}, \apjl, 818, L19

\bibitem[{{Nomoto}, {Kobayashi} \& {Tominaga}(2013){Nomoto}, {Kobayashi}, \&
  {Tominaga}}]{2013ARA&A..51..457N}
{Nomoto} K., {Kobayashi} C., {Tominaga} N., 2013, \araa, 51, 457

\bibitem[{{Nordin} {et~al}\mbox{.}(2011){Nordin}, {{\"O}stman}, {Goobar},
  {Balland}, {Lampeitl}, {Nichol}, {Sako}, {Schneider}, {Smith}, {Sollerman},
  \& {Wheeler}}]{2011ApJ...734...42N}
{Nordin} J. {et~al.}, 2011, \apj, 734, 42

\bibitem[{{O'Donnell}(1994)}]{1994ApJ...422..158O}
{O'Donnell} J.~E., 1994, \apj, 422, 158

\bibitem[{{Osterbrock} \& {Ferland}(2006)}]{2006agna.book.....O}
{Osterbrock} D.~E., {Ferland} G.~J., 2006, {Astrophysics of gaseous nebulae and
  active galactic nuclei}

\bibitem[{{Pan} {et~al}\mbox{.}(2014){Pan}, {Sullivan}, {Maguire}, {Hook},
  {Nugent}, {Howell}, {Arcavi}, {Botyanszki}, {Cenko}, {DeRose}, {Fakhouri},
  {Gal-Yam}, {Hsiao}, {Kulkarni}, {Laher}, {Lidman}, {Nordin}, {Walker}, \&
  {Xu}}]{2014MNRAS.438.1391P}
{Pan} Y.-C. {et~al.}, 2014, \mnras, 438, 1391

\bibitem[{{Peimbert} \& {Costero}(1969)}]{1969BOTT....5....3P}
{Peimbert} M., {Costero} R., 1969, Boletin de los Observatorios Tonantzintla y
  Tacubaya, 5, 3

\bibitem[{{P{\'e}rez-Montero} \& {Contini}(2009)}]{2009MNRAS.398..949P}
{P{\'e}rez-Montero} E., {Contini} T., 2009, \mnras, 398, 949

\bibitem[{{Pettini} \& {Pagel}(2004)}]{2004MNRAS.348L..59P}
{Pettini} M., {Pagel} B.~E.~J., 2004, \mnras, 348, L59

\bibitem[{{Phillips}(1993)}]{1993ApJ...413L.105P}
{Phillips} M.~M., 1993, \apjl, 413, L105

\bibitem[{{Phillips} {et~al}\mbox{.}(1999){Phillips}, {Lira}, {Suntzeff},
  {Schommer}, {Hamuy}, \& {Maza}}]{1999AJ....118.1766P}
{Phillips} M.~M., {Lira} P., {Suntzeff} N.~B., {Schommer} R.~A., {Hamuy} M.,
  {Maza} J., 1999, \aj, 118, 1766

\bibitem[{{Pilyugin}(2001{\natexlab{a}})}]{2001A&A...369..594P}
{Pilyugin} L.~S., 2001{\natexlab{a}}, \aap, 369, 594

\bibitem[{{Pilyugin}(2001{\natexlab{b}})}]{2001A&A...374..412P}
{Pilyugin} L.~S., 2001{\natexlab{b}}, \aap, 374, 412

\bibitem[{{Pilyugin} \& {Thuan}(2005)}]{2005ApJ...631..231P}
{Pilyugin} L.~S., {Thuan} T.~X., 2005, \apj, 631, 231

\bibitem[{{Prieto}, {Rest} \& {Suntzeff}(2006){Prieto}, {Rest}, \&
  {Suntzeff}}]{2006ApJ...647..501P}
{Prieto} J.~L., {Rest} A., {Suntzeff} N.~B., 2006, \apj, 647, 501

\bibitem[{{Riess}, {Press} \& {Kirshner}(1996){Riess}, {Press}, \&
  {Kirshner}}]{1996ApJ...473...88R}
{Riess} A.~G., {Press} W.~H., {Kirshner} R.~P., 1996, \apj, 473, 88

\bibitem[{{Rigault} {et~al}\mbox{.}(2013){Rigault}, {Copin}, {Aldering},
  {Antilogus}, {Aragon}, {Bailey}, {Baltay}, {Bongard}, {Buton}, {Canto},
  {Cellier-Holzem}, {Childress}, {Chotard}, {Fakhouri}, {Feindt}, {Fleury},
  {Gangler}, {Greskovic}, {Guy}, {Kim}, {Kowalski}, {Lombardo}, {Nordin},
  {Nugent}, {Pain}, {P{\'e}contal}, {Pereira}, {Perlmutter}, {Rabinowitz},
  {Runge}, {Saunders}, {Scalzo}, {Smadja}, {Tao}, {Thomas}, \&
  {Weaver}}]{2013A&A...560A..66R}
{Rigault} M. {et~al.}, 2013, \aap, 560, A66

\bibitem[{{Sako} {et~al}\mbox{.}(2014){Sako}, {Bassett}, {Becker}, {Brown},
  {Campbell}, {Cane}, {Cinabro}, {D'Andrea}, {Dawson}, {DeJongh}, {Depoy},
  {Dilday}, {Doi}, {Filippenko}, {Fischer}, {Foley}, {Frieman}, {Galbany},
  {Garnavich}, {Goobar}, {Gupta}, {Hill}, {Hayden}, {Hlozek}, {Holtzman},
  {Hopp}, {Jha}, {Kessler}, {Kollatschny}, {Leloudas}, {Marriner}, {Marshall},
  {Miquel}, {Morokuma}, {Mosher}, {Nichol}, {Nordin}, {Olmstead}, {Ostman},
  {Prieto}, {Richmond}, {Romani}, {Sollerman}, {Stritzinger}, {Schneider},
  {Smith}, {Wheeler}, {Yasuda}, \& {Zheng}}]{2014arXiv1401.3317S}
{Sako} M. {et~al.}, 2014, ArXiv e-prints

\bibitem[{{S{\'a}nchez} {et~al}\mbox{.}(2014){S{\'a}nchez}, {Rosales-Ortega},
  {Iglesias-P{\'a}ramo}, {Moll{\'a}}, {Barrera-Ballesteros}, {Marino},
  {P{\'e}rez}, {S{\'a}nchez-Blazquez}, {Gonz{\'a}lez Delgado}, {Cid Fernandes},
  {de Lorenzo-C{\'a}ceres}, {Mendez-Abreu}, {Galbany}, {Falcon-Barroso},
  {Miralles-Caballero}, {Husemann}, {Garc{\'{\i}}a-Benito}, {Mast}, {Walcher},
  {Gil de Paz}, {Garc{\'{\i}}a-Lorenzo}, {Jungwiert}, {V{\'{\i}}lchez},
  {J{\'{\i}}lkov{\'a}}, {Lyubenova}, {Cortijo-Ferrero}, {D{\'{\i}}az},
  {Wisotzki}, {M{\'a}rquez}, {Bland-Hawthorn}, {Ellis}, {van de Ven}, {Jahnke},
  {Papaderos}, {Gomes}, {Mendoza}, \&
  {L{\'o}pez-S{\'a}nchez}}]{2014A&A...563A..49S}
{S{\'a}nchez} S.~F. {et~al.}, 2014, \aap, 563, A49

\bibitem[{{S{\'a}nchez} {et~al}\mbox{.}(2013){S{\'a}nchez}, {Rosales-Ortega},
  {Jungwiert}, {Iglesias-P{\'a}ramo}, {V{\'{\i}}lchez}, {Marino}, {Walcher},
  {Husemann}, {Mast}, {Monreal-Ibero}, {Cid Fernandes}, {P{\'e}rez},
  {Gonz{\'a}lez Delgado}, {Garc{\'{\i}}a-Benito}, {Galbany}, {van de Ven},
  {Jahnke}, {Flores}, {Bland-Hawthorn}, {L{\'o}pez-S{\'a}nchez}, {Stanishev},
  {Miralles-Caballero}, {D{\'{\i}}az}, {S{\'a}nchez-Blazquez}, {Moll{\'a}},
  {Gallazzi}, {Papaderos}, {Gomes}, {Gruel}, {P{\'e}rez}, {Ruiz-Lara},
  {Florido}, {de Lorenzo-C{\'a}ceres}, {Mendez-Abreu}, {Kehrig}, {Roth},
  {Ziegler}, {Alves}, {Wisotzki}, {Kupko}, {Quirrenbach}, {Bomans}, \& {Califa
  Collaboration}}]{2013A&A...554A..58S}
{S{\'a}nchez} S.~F. {et~al.}, 2013, \aap, 554, A58

\bibitem[{{S{\'a}nchez-Bl{\'a}zquez}
  {et~al}\mbox{.}(2006){S{\'a}nchez-Bl{\'a}zquez}, {Peletier},
  {Jim{\'e}nez-Vicente}, {Cardiel}, {Cenarro}, {Falc{\'o}n-Barroso}, {Gorgas},
  {Selam}, \& {Vazdekis}}]{2006MNRAS.371..703S}
{S{\'a}nchez-Bl{\'a}zquez} P. {et~al.}, 2006, \mnras, 371, 703

\bibitem[{{Schlafly} \& {Finkbeiner}(2011)}]{2011ApJ...737..103S}
{Schlafly} E.~F., {Finkbeiner} D.~P., 2011, \apj, 737, 103

\bibitem[{{Scolnic} {et~al}\mbox{.}(2017){Scolnic}, {Jones}, {Rest}, {Pan},
  {Chornock}, {Foley}, {Huber}, {Kessler}, {Narayan}, {Riess}, {Rodney},
  {Berger}, {Challis}, {Drout}, {Finkbeiner}, {Lunnan}, {Kirshner}, {Sanders},
  {Schlafly}, {Smartt}, {Stubbs}, {Tonry}, {Wood-Vasey}, {Foley}, {Hand},
  {Johnson}, {Burgett}, {Chambers}, {Draper}, {Hodapp}, {Kaiser}, {Kudritzki},
  {Magnier}, {Metcalfe}, {Bresolin}, {Gall}, {Kotak}, {McCrum}, \&
  {Smith}}]{2017arXiv171000845S}
{Scolnic} D.~M. {et~al.}, 2017, ArXiv e-prints

\bibitem[{{Shaw} \& {Dufour}(1995)}]{1995PASP..107..896S}
{Shaw} R.~A., {Dufour} R.~J., 1995, \pasp, 107, 896

\bibitem[{{Stanishev} {et~al}\mbox{.}(2012){Stanishev}, {Rodrigues},
  {Mour{\~a}o}, \& {Flores}}]{2012A&A...545A..58S}
{Stanishev} V., {Rodrigues} M., {Mour{\~a}o} A., {Flores} H., 2012, \aap, 545,
  A58

\bibitem[{{Stasi{\'n}ska}(2010)}]{2010IAUS..262...93S}
{Stasi{\'n}ska} G., 2010, in IAU Symposium, Vol. 262, Stellar Populations -
  Planning for the Next Decade, {Bruzual} G.~R., {Charlot} S., eds., pp. 93--96

\bibitem[{{Storey} \& {Hummer}(1995)}]{1995MNRAS.272...41S}
{Storey} P.~J., {Hummer} D.~G., 1995, \mnras, 272, 41

\bibitem[{{Stoughton} {et~al}\mbox{.}(2002){Stoughton}, {Lupton}, {Bernardi},
  {Blanton}, {Burles}, {Castander}, {Connolly}, {Eisenstein}, {Frieman},
  {Hennessy}, {Hindsley}, {Ivezi{\'c}}, {Kent}, {Kunszt}, {Lee}, {Meiksin},
  {Munn}, {Newberg}, {Nichol}, {Nicinski}, {Pier}, {Richards}, {Richmond},
  {Schlegel}, {Smith}, {Strauss}, {SubbaRao}, {Szalay}, {Thakar}, {Tucker},
  {Vanden Berk}, {Yanny}, {Adelman}, {Anderson}, {Anderson}, {Annis},
  {Bahcall}, {Bakken}, {Bartelmann}, {Bastian}, {Bauer}, {Berman},
  {B{\"o}hringer}, {Boroski}, {Bracker}, {Briegel}, {Briggs}, {Brinkmann},
  {Brunner}, {Carey}, {Carr}, {Chen}, {Christian}, {Colestock}, {Crocker},
  {Csabai}, {Czarapata}, {Dalcanton}, {Davidsen}, {Davis}, {Dehnen},
  {Dodelson}, {Doi}, {Dombeck}, {Donahue}, {Ellman}, {Elms}, {Evans}, {Eyer},
  {Fan}, {Federwitz}, {Friedman}, {Fukugita}, {Gal}, {Gillespie}, {Glazebrook},
  {Gray}, {Grebel}, {Greenawalt}, {Greene}, {Gunn}, {de Haas}, {Haiman},
  {Haldeman}, {Hall}, {Hamabe}, {Hansen}, {Harris}, {Harris}, {Harvanek},
  {Hawley}, {Hayes}, {Heckman}, {Helmi}, {Henden}, {Hogan}, {Hogg}, {Holmgren},
  {Holtzman}, {Huang}, {Hull}, {Ichikawa}, {Ichikawa}, {Johnston}, {Kauffmann},
  {Kim}, {Kimball}, {Kinney}, {Klaene}, {Kleinman}, {Klypin}, {Knapp},
  {Korienek}, {Krolik}, {Kron}, {Krzesi{\'n}ski}, {Lamb}, {Leger},
  {Limmongkol}, {Lindenmeyer}, {Long}, {Loomis}, {Loveday}, {MacKinnon},
  {Mannery}, {Mantsch}, {Margon}, {McGehee}, {McKay}, {McLean}, {Menou},
  {Merelli}, {Mo}, {Monet}, {Nakamura}, {Narayanan}, {Nash}, {Neilsen},
  {Newman}, {Nitta}, {Odenkirchen}, {Okada}, {Okamura}, {Ostriker}, {Owen},
  {Pauls}, {Peoples}, {Peterson}, {Petravick}, {Pope}, {Pordes}, {Postman},
  {Prosapio}, {Quinn}, {Rechenmacher}, {Rivetta}, {Rix}, {Rockosi}, {Rosner},
  {Ruthmansdorfer}, {Sandford}, {Schneider}, {Scranton}, {Sekiguchi}, {Sergey},
  {Sheth}, {Shimasaku}, {Smee}, {Snedden}, {Stebbins}, {Stubbs}, {Szapudi},
  {Szkody}, {Szokoly}, {Tabachnik}, {Tsvetanov}, {Uomoto}, {Vogeley}, {Voges},
  {Waddell}, {Walterbos}, {Wang}, {Watanabe}, {Weinberg}, {White}, {White},
  {Wilhite}, {Wolfe}, {Yasuda}, {York}, {Zehavi}, \&
  {Zheng}}]{2002AJ....123..485S}
{Stoughton} C. {et~al.}, 2002, \aj, 123, 485

\bibitem[{{Sullivan} {et~al}\mbox{.}(2010){Sullivan}, {Conley}, {Howell},
  {Neill}, {Astier}, {Balland}, {Basa}, {Carlberg}, {Fouchez}, {Guy}, {Hardin},
  {Hook}, {Pain}, {Palanque-Delabrouille}, {Perrett}, {Pritchet}, {Regnault},
  {Rich}, {Ruhlmann-Kleider}, {Baumont}, {Hsiao}, {Kronborg}, {Lidman},
  {Perlmutter}, \& {Walker}}]{2010MNRAS.406..782S}
{Sullivan} M. {et~al.}, 2010, \mnras, 406, 782

\bibitem[{{Sullivan} {et~al}\mbox{.}(2011){Sullivan}, {Guy}, {Conley},
  {Regnault}, {Astier}, {Balland}, {Basa}, {Carlberg}, {Fouchez}, {Hardin},
  {Hook}, {Howell}, {Pain}, {Palanque-Delabrouille}, {Perrett}, {Pritchet},
  {Rich}, {Ruhlmann-Kleider}, {Balam}, {Baumont}, {Ellis}, {Fabbro},
  {Fakhouri}, {Fourmanoit}, {Gonz{\'a}lez-Gait{\'a}n}, {Graham}, {Hudson},
  {Hsiao}, {Kronborg}, {Lidman}, {Mourao}, {Neill}, {Perlmutter}, {Ripoche},
  {Suzuki}, \& {Walker}}]{2011ApJ...737..102S}
{Sullivan} M. {et~al.}, 2011, \apj, 737, 102

\bibitem[{{Sullivan} {et~al}\mbox{.}(2006){Sullivan}, {Le Borgne}, {Pritchet},
  {Hodsman}, {Neill}, {Howell}, {Carlberg}, {Astier}, {Aubourg}, {Balam},
  {Basa}, {Conley}, {Fabbro}, {Fouchez}, {Guy}, {Hook}, {Pain},
  {Palanque-Delabrouille}, {Perrett}, {Regnault}, {Rich}, {Taillet}, {Baumont},
  {Bronder}, {Ellis}, {Filiol}, {Lusset}, {Perlmutter}, {Ripoche}, \&
  {Tao}}]{2006ApJ...648..868S}
{Sullivan} M. {et~al.}, 2006, \apj, 648, 868

\bibitem[{{Suzuki} {et~al}\mbox{.}(2012){Suzuki}, {Rubin}, {Lidman},
  {Aldering}, {Amanullah}, {Barbary}, {Barrientos}, {Botyanszki}, {Brodwin},
  {Connolly}, {Dawson}, {Dey}, {Doi}, {Donahue}, {Deustua}, {Eisenhardt},
  {Ellingson}, {Faccioli}, {Fadeyev}, {Fakhouri}, {Fruchter}, {Gilbank},
  {Gladders}, {Goldhaber}, {Gonzalez}, {Goobar}, {Gude}, {Hattori}, {Hoekstra},
  {Hsiao}, {Huang}, {Ihara}, {Jee}, {Johnston}, {Kashikawa}, {Koester},
  {Konishi}, {Kowalski}, {Linder}, {Lubin}, {Melbourne}, {Meyers}, {Morokuma},
  {Munshi}, {Mullis}, {Oda}, {Panagia}, {Perlmutter}, {Postman}, {Pritchard},
  {Rhodes}, {Ripoche}, {Rosati}, {Schlegel}, {Spadafora}, {Stanford},
  {Stanishev}, {Stern}, {Strovink}, {Takanashi}, {Tokita}, {Wagner}, {Wang},
  {Yasuda}, {Yee}, \& {Supernova Cosmology Project}}]{2012ApJ...746...85S}
{Suzuki} N. {et~al.}, 2012, \apj, 746, 85

\bibitem[{{Thomas} {et~al}\mbox{.}(2013){Thomas}, {Steele}, {Maraston},
  {Johansson}, {Beifiori}, {Pforr}, {Str{\"o}mb{\"a}ck}, {Tremonti}, {Wake},
  {Bizyaev}, {Bolton}, {Brewington}, {Brownstein}, {Comparat}, {Kneib},
  {Malanushenko}, {Malanushenko}, {Oravetz}, {Pan}, {Parejko}, {Schneider},
  {Shelden}, {Simmons}, {Snedden}, {Tanaka}, {Weaver}, \&
  {Yan}}]{2013MNRAS.431.1383T}
{Thomas} D. {et~al.}, 2013, \mnras, 431, 1383

\bibitem[{{Timmes}, {Brown} \& {Truran}(2003){Timmes}, {Brown}, \&
  {Truran}}]{2003ApJ...590L..83T}
{Timmes} F.~X., {Brown} E.~F., {Truran} J.~W., 2003, \apjl, 590, L83

\bibitem[{{Tremonti} {et~al}\mbox{.}(2004){Tremonti}, {Heckman}, {Kauffmann},
  {Brinchmann}, {Charlot}, {White}, {Seibert}, {Peng}, {Schlegel}, {Uomoto},
  {Fukugita}, \& {Brinkmann}}]{2004ApJ...613..898T}
{Tremonti} C.~A. {et~al.}, 2004, \apj, 613, 898

\bibitem[{{Veilleux} \& {Osterbrock}(1987)}]{1987ApJS...63..295V}
{Veilleux} S., {Osterbrock} D.~E., 1987, \apjs, 63, 295

\bibitem[{{Wolf} {et~al}\mbox{.}(2016){Wolf}, {D'Andrea}, {Gupta}, {Sako},
  {Fischer}, {Kessler}, {Jha}, {March}, {Scolnic}, {Fischer}, {Campbell},
  {Nichol}, {Olmstead}, {Richmond}, {Schneider}, \&
  {Smith}}]{2016ApJ...821..115W}
{Wolf} R.~C. {et~al.}, 2016, \apj, 821, 115

\bibitem[{{York} {et~al}\mbox{.}(2000){York}, {Adelman}, {Anderson},
  {Anderson}, {Annis}, {Bahcall}, {Bakken}, {Barkhouser}, {Bastian}, {Berman},
  {Boroski}, {Bracker}, {Briegel}, {Briggs}, {Brinkmann}, {Brunner}, {Burles},
  {Carey}, {Carr}, {Castander}, {Chen}, {Colestock}, {Connolly}, {Crocker},
  {Csabai}, {Czarapata}, {Davis}, {Doi}, {Dombeck}, {Eisenstein}, {Ellman},
  {Elms}, {Evans}, {Fan}, {Federwitz}, {Fiscelli}, {Friedman}, {Frieman},
  {Fukugita}, {Gillespie}, {Gunn}, {Gurbani}, {de Haas}, {Haldeman}, {Harris},
  {Hayes}, {Heckman}, {Hennessy}, {Hindsley}, {Holm}, {Holmgren}, {Huang},
  {Hull}, {Husby}, {Ichikawa}, {Ichikawa}, {Ivezi{\'c}}, {Kent}, {Kim},
  {Kinney}, {Klaene}, {Kleinman}, {Kleinman}, {Knapp}, {Korienek}, {Kron},
  {Kunszt}, {Lamb}, {Lee}, {Leger}, {Limmongkol}, {Lindenmeyer}, {Long},
  {Loomis}, {Loveday}, {Lucinio}, {Lupton}, {MacKinnon}, {Mannery}, {Mantsch},
  {Margon}, {McGehee}, {McKay}, {Meiksin}, {Merelli}, {Monet}, {Munn},
  {Narayanan}, {Nash}, {Neilsen}, {Neswold}, {Newberg}, {Nichol}, {Nicinski},
  {Nonino}, {Okada}, {Okamura}, {Ostriker}, {Owen}, {Pauls}, {Peoples},
  {Peterson}, {Petravick}, {Pier}, {Pope}, {Pordes}, {Prosapio},
  {Rechenmacher}, {Quinn}, {Richards}, {Richmond}, {Rivetta}, {Rockosi},
  {Ruthmansdorfer}, {Sandford}, {Schlegel}, {Schneider}, {Sekiguchi}, {Sergey},
  {Shimasaku}, {Siegmund}, {Smee}, {Smith}, {Snedden}, {Stone}, {Stoughton},
  {Strauss}, {Stubbs}, {SubbaRao}, {Szalay}, {Szapudi}, {Szokoly}, {Thakar},
  {Tremonti}, {Tucker}, {Uomoto}, {Vanden Berk}, {Vogeley}, {Waddell}, {Wang},
  {Watanabe}, {Weinberg}, {Yanny}, {Yasuda}, \& {SDSS
  Collaboration}}]{2000AJ....120.1579Y}
{York} D.~G. {et~al.}, 2000, \aj, 120, 1579

\end{thebibliography}

\end{document}